\documentclass[fleqn,usenatbib]{mnras}

\usepackage[T1]{fontenc}
\usepackage{ae,aecompl}

\usepackage{graphicx}	% Including figure files
\usepackage{amsmath}	% Advanced maths commands
\usepackage{amssymb}	% Extra maths symbols
\usepackage{lipsum}
\usepackage{multirow}
\usepackage{multicol}
\usepackage{xcolor}    % for colored text
\usepackage{subcaption}

\newcommand{\mrm}{\mathrm}
\newcommand{\mbf}{\mathbf}

\newcommand{\fNL}{$f_{\mathrm{NL}} \ $}
\newcommand{\fNLloc}{$f_{\mathrm{NL}}^\mathrm{local} \ $}

\title[]{The clustering of galaxies in the completed SDSS-IV extended  Baryon Oscillation Spectroscopic Survey: Primordial non-Gaussianity in Fourier Space}

\author[E.-M. Mueller et al.]{Eva-Maria Mueller$^{1,9}$\thanks{E-mail: eva.mueller@physics.ox.ac.uk},
Mehdi Rezaie${^2}$,
Will J. Percival$^{3,4,5}$,
Ashley J. Ross$^{6}$,
\newauthor
Rossana Ruggeri$^{7}$,
Hee-Jong Seo$^{2}$,
H\'ector Gil-Mar\'in$^{8}$,
Julian Bautista$^{9,10}$,
\newauthor
Joel R. Brownstein$^{11}$,
Kyle Dawson$^{11}$,
Axel de la Macorra$^{12}$,
\newauthor
Nathalie Palanque-Delabrouille$^{13}$,
Graziano Rossi$^{14}$,
Donald P. Schneider$^{15,16}$, 
\newauthor
Christophe Yèche$^{13}$
\\
\scriptsize $^{1}$ Department of Physics, University of Oxford, Denys Wilkinson Building, Keble Road, Oxford OX1 3RH, UK\\
\scriptsize $^{2}$ Department of Physics and Astronomy, Ohio University, 251B Clippinger Labs, Athens, OH 45701, USA\vspace*{-2pt} \\
\scriptsize $^{3}$ Waterloo Centre for Astrophysics, University of Waterloo, 200 University Ave W, Waterloo, ON N2L 3G1, Canada \\
\scriptsize $^{4}$ Department of Physics and Astronomy, University of Waterloo, 200 University Ave W, Waterloo, ON N2L 3G1, Canada \\
\scriptsize $^{5}$ Perimeter Institute for Theoretical Physics, 31 Caroline St. North, Waterloo, ON N2L 2Y5, Canada \\
\scriptsize $^{6}$ Center for Cosmology and Astro-Particle Physics, Ohio State University, Columbus, Ohio, USA \\
\scriptsize $^{7}$ School of Mathematics and Physics, University of Queensland, Brisbane, QLD 4072, Australia \\
\scriptsize $^{8}$ICC, University of Barcelona, IEEC-UB, Mart\'i i Franqu\`es, 1, E-08028 Barcelona, Spain \\
\scriptsize $^{9}$  Institute of Cosmology and Gravitation, Dennis Sciama Building, University of Portsmouth, Portsmouth PO1 3FX, UK\\
\scriptsize $^{10}$ Aix Marseille University, CNRS/IN2P3, CPPM, Marseille, France \\
\scriptsize $^{11}$ Department of Physics and Astronomy, University of Utah, 115 S. 1400 E., Salt Lake City, UT 84112, USA \\
\scriptsize $^{12}$Instituto de Física, Universidad Nacional Autónoma de México, Apdo. Postal 20-364, México \\
\scriptsize $^{13}$IRFU, CEA, Centre d’Etudes Saclay, 91191 Gif-Sur-Yvette Cedex, France \\
\scriptsize $^{14}$ Department of Astronomy and Space Science, Sejong University, 209, Neungdong-ro, Gwangjin-gu, Seoul, South Korea\\
\scriptsize $^{15}$Department of Astronomy and Astrophysics, The Pennsylvania State University, University Park, PA 16802 \\
\scriptsize $^{16}$Institute for Gravitation and the Cosmos, The Pennsylvania State University, University Park, PA 16802 \\
}

\date{Accepted XXX. Received YYY; in original form ZZZ}

\pubyear{2021}
\begin{document}
\label{firstpage}
\pagerange{\pageref{firstpage}--\pageref{lastpage}}
\maketitle

%%%%%%%%%%%%%%%%%%% ABSTRACT %%%%%%%%%%%%%%%%%%%
% Abstract of the paper
\begin{abstract}
We present measurements of the local primordial non-Gaussianity parameter \fNLloc from the clustering of 343,708 quasars with redshifts 0.8 < z < 2.2 distributed over 4808 square degrees from the final data release (DR16) of the extended Baryon acoustic Oscillation Spectroscopic Survey (eBOSS), the largest volume spectroscopic survey up to date. Our analysis is performed in Fourier space, using the power spectrum monopole at very large scales to constrain the scale dependent halo bias. We carefully assess the impact of systematics on our measurement and test multiple contamination removal methods. We demonstrate the robustness of our analysis pipeline with EZ-mock catalogues that simulate the eBOSS DR16 target selection. We find $f_\mrm{NL}=-12\pm 21$ (68\% confidence) for the main clustering sample including quasars with redshifts between 0.8 and 2.2, after exploiting a novel neural network scheme for cleaning the DR16 sample and in particular after applying redshift weighting techniques, designed for non-Gaussianity measurement from large scales structure, to optimize our analysis, which improve our results by 37\%.
\end{abstract}

% Select between one and six entries from the list of approved keywords.
% Don't make up new ones.

%%%%%%%%%%%%%%%%%%%%%%%%%%%%%%%%%%%%%%%%%%%%%%%%%%

%%%%%%%%%%%%%%%%% BODY OF PAPER %%%%%%%%%%%%%%%%%%

%%%%%%%%%%%%%%%%%%%%%%%%%%%%%%%%%%%%%%%%%%%%%%%%%%%%
%%%%%%%%%%%%%%%%%%% INTRODUCTION %%%%%%%%%%%%%%%%%%%
%%%%%%%%%%%%%%%%%%%%%%%%%%%%%%%%%%%%%%%%%%%%%%%%%%%%

\section{Introduction}

The aim of this paper is to constrain primordial non-Gaussianity (PNG) using the large scale structure (LSS) in the clustering of quasars covering the largest volume of the Universe surveyed to date. We use the quasars within the final data release (DR16; \citealt{SDSS-DR16}) of the extended Baryon Oscillation Spectroscopic Survey (eBOSS; \citealt{dawson_sdss-iv_2016}), which is a part of the Sloan Digital Sky Survey (SDSS; \citealt{blanton_sloan_2017}) program. 

PNG arise during inflation, a phase of rapid expansion in the very early Universe, and can be used to differentiate between different underlying physical models. Slow roll, single field inflation generates a primordial gravitational potential that can be well approximated by a Gaussian random field \citep{bardeen_statistics_1986}. Alternative inflationary models (e.g., multiple field) predict there to be significant non-Gaussian components to the potential \citep{Bartolo04}. In this paper, we focus on the simplest case of PNG, assuming local scale-independent PNG parameterized by $f_\mathrm{NL}$. The tightest constraints on this form of PNG are currently from the Planck satellite \citep{Planck2018-fnl} measuring the cosmic microwave background (CMB), who found $f_\mathrm{NL}=-0.9\pm5.1$. However, the LSS in the clustering of galaxies or quasars offers an alternative pathway to detect PNG that has the potential to improve upon these constraints in the future \citep{Dore-spherex-cosmo,Euclid-SKA-fnl,Radio-fnl,MSE-fnl}.

One effect of PNG is to add a scale-dependence to the large scale linear bias of galaxies with respect to the underlying dark matter (\citealt{dalal_imprints_2008,Matarrese08,Slosar08}, and discussed further in Section~\ref{sec:model}). The unique $f_\mathrm{NL}/k^2$ dependence of the galaxy bias introduced by PNG means that the signal increases to large scales, and requires a large volume survey with exquisite control of systematic errors. For any sample of galaxies, the bias and hence the PNG signal will change with redshift. Weights were introduced by \citet{Mueller19} to optimally include all of the information across redshift and are used in our paper. Including constraints from the bispectrum and trispectrum would further improve these predictions \citep{Karagiannis18,gualdi2021joint}, but we do not do this because of the increased complexity of such an analysis. 

Previous measurements of PNG using LSS include \citet{Slosar08,Ross13-fnl,Giannantonio14-fnl,Ho15-fnl,Leistedt14-fnl,eboss-dr14-fnl}. In general these studies are systematic limited, with variations in observed large-scale galaxy density driven by such effects as extinction, depth, air-mass or star density. Because the clustering only needs to be measured on very large-scales, the use of photometric redshifts does not significantly degrade the statistical error. However, our ability to understand and correct for problems in the data is significantly improved using spectroscopic redshifts. 

Quasars offer an ideal large-scale structure tracer for measuring PNG because they are highly biased \citep{Laurent-dr14-qso} and their brightness means they can be observed over large volumes. The eBOSS quasar sample we use \citep{ebossDR16catalogue}, includes objects up to a redshift of 2.2 over 4808\,deg$^2$, with a physical volume of 68.7\,Gpc$^3$ (assuming our baseline $\Lambda$CDM model), and is the largest LSS sample in volume to date. Further details are provided in Section~\ref{sec:cat}. This is a subset of the full sample of quasars observed by SDSS, which is described in \citet{Lyke-dr16-qsocat}, and is subsampled in order to provide a catalogue for clustering measurements for which we can model the selection function. The large volume covered by the quasar catalogue is crucial for the detection of local PNG through its effect on the large-scale bias. Hence, while the eBOSS quasar sample was primarily designed \citep{zhao_extended_2016} for the measurements of Baryon Acoustic Oscillations (BAO) and Redshift-Space Distortions (RSD), it is also ideal to study PNG.

\cite{eboss-dr14-fnl} measured $f_\mathrm{NL}$ using the eBOSS DR14 quasar sample, finding $-51<f_\mathrm{NL}<21$ at 95 \% confidence. Their sample was taken when eBOSS was part-complete covered $2114$ square degrees, about half of the DR16 sample we analyse. Our analysis also differs in that we present a novel neural network based method to remove systematic density fluctuations in this sample \citep{RezaiCompanion}, which is summarised in Section~\ref{sec:sys}. Standard BAO and RSD measurements made with this sample are presented in \citet{Hou-eboss-qso-xi,Neveux-eboss-qso-pk}, together with a blind test of their methodology in \citet{Smith-eboss-qso-mocks}. At redshifts $z>2.1$, eBOSS also made Lyman-$\alpha$ forest measurements using a high redshift quasar sample \citep{eBOSS-DR16-Lya}. The BAO measurement using the same neural network catalog we implement in this paper is presented in \citet{Merz2021}.

The eBOSS DR16 quasar sample \citep{Lyke-dr16-qsocat,ebossDR16catalogue} is the largest sample of spectroscopically confirmed quasars available to date, and is one of three target classes observed within eBOSS. Samples of Emission line galaxies  \citep{raichoor20a,tamone20a,demattia20a} and Luminous Red Galaxies \citep{LRG_corr,gil-marin20a} were also observed and used to make cosmological inferences \citep{eBOSS_Cosmology}.

As we are interested in very large scales, we have to be careful when calculating the power spectrum and allowing for the window function and integral constraint: our methods are described in Section~\ref{sec:method} and tested using mock catalogues (described in Section~\ref{sec:mocks}) as described in Section~\ref{sec:tests}. We do not include relativistic effects which, although they are degenerate with  $f_\mathrm{NL}$, are not expected to have an influence larger than $f_\mathrm{NL}\pm1$ on the scales and redshifts fitted \citep{wang20}. Similarly, at the level of constraints that can be achieved with eBOSS data, we do not allow for the non-Gaussian nature of the likelihood as discussed, for example by \cite{Kalus19, wang19}. The latter assumption is validated by our analyses of mock catalogues. We present the results from our analysis in Section~\ref{sec:results} and discuss them in Section~\ref{sec:conclusion}.

%%%%%%%%%%%%%%%%%%%%%%%%%%%%%%%%%%%%%%%%%%%%%%%%%%
%%%%%%%%%%%%%%%%%%% PHYSICAL MODEL %%%%%%%%%%%%%%%
%%%%%%%%%%%%%%%%%%%%%%%%%%%%%%%%%%%%%%%%%%%%%%%%%%

\section{Physical Model}  \label{sec:model}

In this Section we provide a brief summary of the mechanism by which local non-Gaussianity causes scale dependent halo bias. Assuming a type of non-Gaussianity that only depends on the local value of the potential, the primordial potential can be parametrised  \citep{1994ApJ...430..447G, 2001PhRvD..63f3002K}
\begin{equation}  \label{eq:phi_ng}
    \Phi = \phi + f_\mathrm{NL} (\phi^2 - \langle \phi^2 \rangle)\,,
\end{equation}
where $\phi$ is a Gaussian random field and $f_\mathrm{NL}$ describes the amplitude of the quadratic non-Gaussian term in the potential. In Fourier space, the potential is related to the density field by $\delta(k)=\alpha(k) \Phi(k)$, with
\begin{equation} \label{eq:alpha}
    \alpha(k) = \frac{2 k^2 T(k) D(z)}{3\Omega_m} \frac{c^2}{H_0^2}         \frac{g(0)}{g(\infty)}\,.
\end{equation}
$T(k)$ is the transfer function, $D(z)$ the linear growth factor normalised with $D(0)=1$. $\Omega_m$ is the matter density and $H_0$ the Hubble constant. The factor $g(\infty)/g(0)$, with $g(z)=(1+z)D(z)$, corrects for the late-time normalisation of $D(z)$, and can be omitted if $D(z)$ is normalised to equal the scale factor during the matter dominated era. Following \citet{Mueller19}, we assume the CMB convention for $f_\mathrm{NL}$ assuming $\Phi$ is the primordial potential. Some authors have previously adopted a "LSS convention" that assumes $\Phi$ is extrapolated to $z=0$, with $f_\mathrm{NL}^\mathrm{LSS}=g(\infty)/g(0) f_\mathrm{NL}^\mathrm{CMB} \approx 1.3  \ f_\mathrm{NL}^\mathrm{CMB}$. 

In the standard ansatz, large-scale halo bias depends on the derivative of the halo mass function with respect to the large-scale matter overdensity field: small changes in the overdensity can give rise to large changes in the density of galaxies as patches of the Universe are pushed over a critical density for collapse \citep{Gunn_Gott_sphericalcollapse}. The non-Gaussian term in Eq.~\ref{eq:phi_ng}, coupled with the $k$-dependence in Eq.~\ref{eq:alpha} thus give rise to a scale dependent halo bias $\Delta b(k)$ \citep{dalal_imprints_2008,Slosar08}
\begin{equation}\label{eq:bias_ng}
  \Delta b(k)= 2(b-p)f_\mathrm{NL} \frac{\delta_\mathrm{crit}}{\alpha(k)}\,,
\end{equation}
where $\delta_\mathrm{crit}=1.686$ and $p$ is a correction due to galaxy selection beyond a Poisson sampling of the halos of a given mass: quasars are thought to result from galaxy mergers, and the merger rate is proportional to $\sigma_8^{-1}$. For quasars, the standard ansatz requires us to determine $\partial\ln n_\mathrm{merger}/\partial\ln\sigma_8$ rather than $\partial\ln n_\mathrm{halo}/\partial\ln\sigma_8$, and this leads to an extra component in $\Delta b$ that changes $p=1$ to $p=1+1/\delta_\mathrm{crit}\simeq1.6$ \citep{Slosar08}. The total bias of quasars, including local non-Gaussianity is modelled as $b_\mathrm{tot} = b + \Delta b(k)$.

%%%%%%%%%%%%%%%%%%%%%%%%%%%%%%%%%%%%%%%%%%%%%%%%%%
%%%%%%%%%%%%%%%%%%% eBOSS QSO dataset %%%%%%%%%%%%
%%%%%%%%%%%%%%%%%%%%%%%%%%%%%%%%%%%%%%%%%%%%%%%%%%
\section{The eBOSS DR16 QSO dataset}
\label{sec:cat}

Our quasar sample is derived from the final data release, (DR16; \citealt{SDSS-DR16}), of the SDSS-IV eBOSS survey. Over 4.5 years, eBOSS used the Sloan Telescope \citep{Gunn:2006tw} and BOSS spectrographs \citep{Smee:2012wd} to obtain spectra of quasar and galaxy targets.

 The algorithms to robustly identify quasars and accurately measure their redshifts from SDSS spectra were developed in \cite{Lyke-dr16-qsocat}, who produced the DR16 quasar catalogue. This quasar catalogue was used as the input to the main clustering sample, used for the BAO and RSD analysis and detailed in \cite{ebossDR16catalogue}. It includes quasars with $0.8<z<2.2$. The catalogues were processed separately in the North and South Galactic Cap (NGC and SGC) footprints. The NGC sample contains 218,209 quasars over 2,924 square degrees and the SGC sample 125,499 quasars over 1,884 square degrees. The sample is simulated with 1000 mock realizations, as described in \cite{Zhao_2021}.

The quasar sample contains strong spurious fluctuations in its projected density on the sky \citep{Myers:2015hpw,Ata:2017dya,ebossDR16catalogue}. These are primarily related to imaging depth and the probabilistic nature of the target selection that is intrinsically dependent on the precision of the photometry; see \cite{Myers:2015hpw} and Section~\ref{sec:sys} for more details. In \cite{ebossDR16catalogue}, these fluctuations were treated by determining weights based on a linear regression including maps of the $g$-band depth, the PSF size in the $i$-band, the sky background in the $i$-band, and the amount of Galactic dust extinction determined by \cite{Schlegel:1997yv}. This level of correction was determined sufficient for the BAO and RSD analyses (\citealt{Hou-eboss-qso-xi, Neveux-eboss-qso-pk, Smith-eboss-qso-mocks}), and is described in more detail in Section~\ref{sec:lin_weight}, but a re-evaluation of the needs for an $f_{\rm NL}$ measurement was promised to be presented here and in \cite{RezaiCompanion}. This is presented in Section~\ref{sec:NN_weights}.

\begin{figure*}
    \centering
    \includegraphics[width=0.95\textwidth]{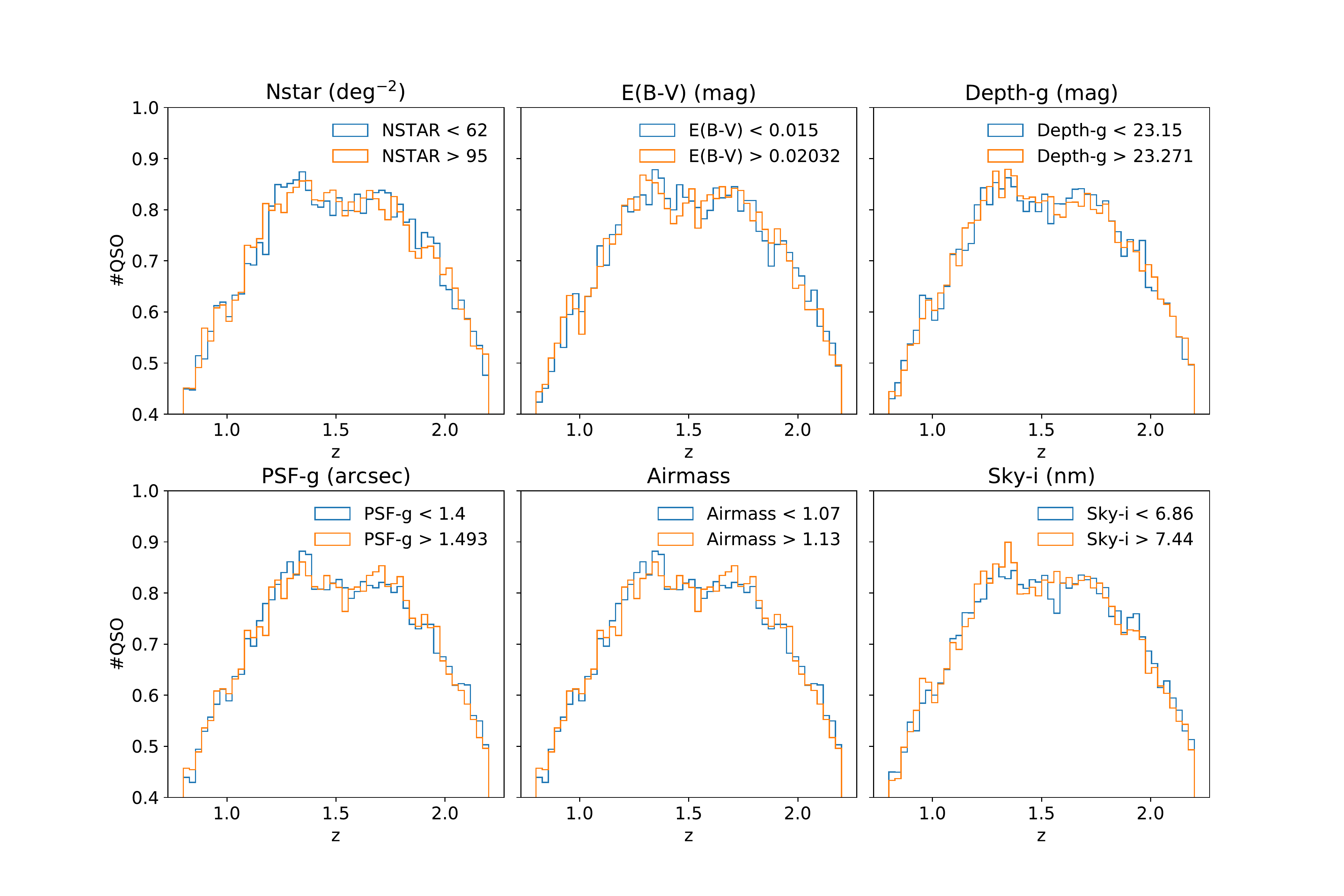}
    \caption{Histograms showing the quasar distribution $n(z)$ for the NGC as a function of redshift for bin of high and low photometric properties. The bins were chosen so that the number of quasars in each bin is equal. Nstar and E(B-V)  refer to the stellar density and extinction respectively, while Depth-g, PSF-g and Sky-i stands for the depth of the imaging data, the seeing and sky background in the g-band and i-band. We find no dependence of $n(z)$ on the photometric systematics, supporting our assertion that we can correct for the systematics with a function that only depends on angular position.}
    \label{fig:nz_histogram}
\end{figure*}

Fortunately, such spurious fluctuations in quasar target density do not couple strongly to changes in the redshift distribution. This can be seen in Fig. \ref{fig:nz_histogram}, where the relative number of quasars as a function redshift is shown when splitting the sample by various imaging properties. The normalized redshift distributions do not appear to depend on the properties of the imaging sample.

%%%%%%%%%%%%%%%%%%%%%%%%%%%%%%%%%%%%%%%%%%%%%%%%%%
%%%%%%%%%%%%%% Photometric Systematics %%%%%%%%%%%
%%%%%%%%%%%%%%%%%%%%%%%%%%%%%%%%%%%%%%%%%%%%%%%%%%
\section{Photometric Systematics}  \label{sec:sys}

The density variations caused by variations in the properties of imaging data typically result in an excess clustering signal on large scales (e.g., $k<0.01~h/{\rm Mpc}$). Unfortunately, the $k^{-2}$-dependent signature of primordial non-Gaussianity is sensitive to the same scales, and thus a robust constraint on the primordial non-Gaussianity parameter $f_{\rm NL}$ demands a thorough and careful assessment of imaging systematics. This reality has been recognized as one of the primary challenges in previous studies of PNG with SDSS data \citep[see, e.g.,][]{Ho15-fnl}.

Systematic effects can be classified in general into two categories: obscuration and contamination. These effects are intermingled too. For instance the effect of Galactic obscuration alters the color of objects and can push some of the quasars out of the color-magnitude selection. Adversely, it might also allow some astrophysical artifacts such as stars to be miss-classified as quasars and enter the selection as contaminants \citep[see, e.g.,][for a discussion]{Crocce2015}.

In what follows, we briefly discuss two map-based regression approaches for the mitigation of imaging systematics. These approaches are computationally efficient, however, their robustness and efficiency in eliminating systematics relies on the available imaging maps and the flexibility of the model. %The trends in the projected density of quasars against imaging properties are essentially different for the North and South Galactic caps due to different targeting efficiencies. Therefore, the treatments are performed for each Galactic cap separately.

%%%%%%%%%%%%%% Linear Weights %%%%%%%%%%%

\subsection{Method I: Linear weights}  \label{sec:lin_weight}

The standard systematics treatment, which is implemented in \cite{Rosss2020MNRAS.498.2354R}, uses a multivariate linear model to capture the response of the observed quasar count map to the templates of imaging properties \citep[e.g.,][]{bautista2018sdss}. The templates include Galactic extinction from \cite{schlegel1998maps}, survey depth in the $g$ band (\textit{depth-g}), sky brightness in the $i$ band (\textit{sky-i}), and seeing in the $i$ band (\textit{psf-i}) from SDSS imaging; and all are pixellated in \textsc{HEALPix} format with \textsc{nside}$=512$. These maps are identified as primary imaging properties which have induced systematic fluctuations in the DR16 quasar sample \cite{Rosss2020MNRAS.498.2354R}. In the standard treatment, the impact of theses templates on the target sample density is assumed to be linear, with four parameters - the amplitude of the effect from each template. The parameters of the linear model are trained by a least-square minimization that reduces the sum of squared error fluctuations in the mean quasar density against imaging properties, including the four primary templates as well as the SDSS stellar density and airmass.

%%%%%%%%%%%%%% Neural Networks %%%%%%%%%%%
\subsection{Method II: Neural Networks}  \label{sec:NN_weights}

As an alternative, we rely on an ensemble of fully connected neural networks (NN) to obtain the systematic weights as described in \citet{RezaiCompanion}. Neural networks are capable of representing complex functions, and thus this new set of systematics weights are able to eliminate nonlinear systematic effects. This is very crucial for cleaning the DR16 sample, especially in the NGC, where we observe a strong nonlinear trend against Galactic extinction. 

We adapt neural networks to predict the projected quasar density given the imaging data properties imprinted on angular maps pixellated using \textsc{HEALPix} \citep{gorski2005ApJ...622..759G} with \textsc{nside}$=512$ as our default resolution, covering the survey. In addition to the four maps used in \cite{Rosss2020MNRAS.498.2354R}, we find that there is a strong dependence on the local stellar density, constructed from Gaia DR2 \citep{gaia2018A&A...616A...1G} with the g-band magnitude cut $12 < {\rm gmag} <17$. We find that the inclusion of the Gaia stellar density was necessary to obtain a small $\chi^{2}$ value consistent with the systematics-free simulations. The importance of this map was only realised after the public released with the catalogue, so it is not included in the default linear weights described in Section~\ref{sec:lin_weight}, which used a stellar template constructed from SDSS. We also confirm that the improvements we see cannot be reproduced with this stellar template. 

A fully connected neural network is constructed from many nonlinear units, called neurons, organized on multiple layers. The information of imaging properties flows forward starting from the first layer through the intermediate layers, and out to the output layer as the predicted quasar density. Each neuron calculates a linear combination of the outputs from the previous layer neurons, and then applies a nonlinear activation function. The arrangement of multiple nonlinear neurons on multiple layers allows the neural network to have the flexibility to model a wide range of functions. Our implementation uses three hidden layers with 20 Rectified Linear Unit neurons on each hidden layer, which is defined as ${\rm ReLU}(x)=\max(x, 0)$. The output layer has a single neuron with a Softplus activation, i.e., Softplus(x)=$\log(1+\exp(x))$, to ensure that the predicted quasar count is positive. The parameters of the neural network are optimized to increase the similarity between the neural network output and the observed projected density field by minimizing the Poisson negative log-likelihood. We incorporate five-fold validation for the training, validation, and testing our neural network. Specifically, we train the parameters with $60\%$ of the data, use $20\%$ of the data to validate and find the best model with minimum validation error. Next, we apply the best model on the remaining $20\%$ of the sample, which is called the test set, and then use the inverse of the predicted quasar count as $w_{\rm systot}$. By permutation of the training, validation, and test sets, we test the methodology on the entire footprint and obtain the systematic weights for the entire sample. Finally, the weights are normalized such that the total weighted number of quasars stays invariant before and after applying the weights.

To test our implementation, we have investigated the impact of changing the resolution of imaging maps, splitting the quasar sample into $0.8<z<1.5$ and $1.5<z<2.2$, and using various combinations of imaging maps. For each of these tests, we characterize the residual systematic error with the $\chi^{2}$ statistics by calculating the angular cross-correlation of the projected quasar density and imaging maps and the fluctuations in the mean density of quasars against imaging properties. We find no evidence for redshift-dependent imaging systematics, and no significant impact when training the networks with templates in \textsc{nside}$=256$. 

\begin{figure*}
    \centering
    \includegraphics[width=0.95\textwidth]{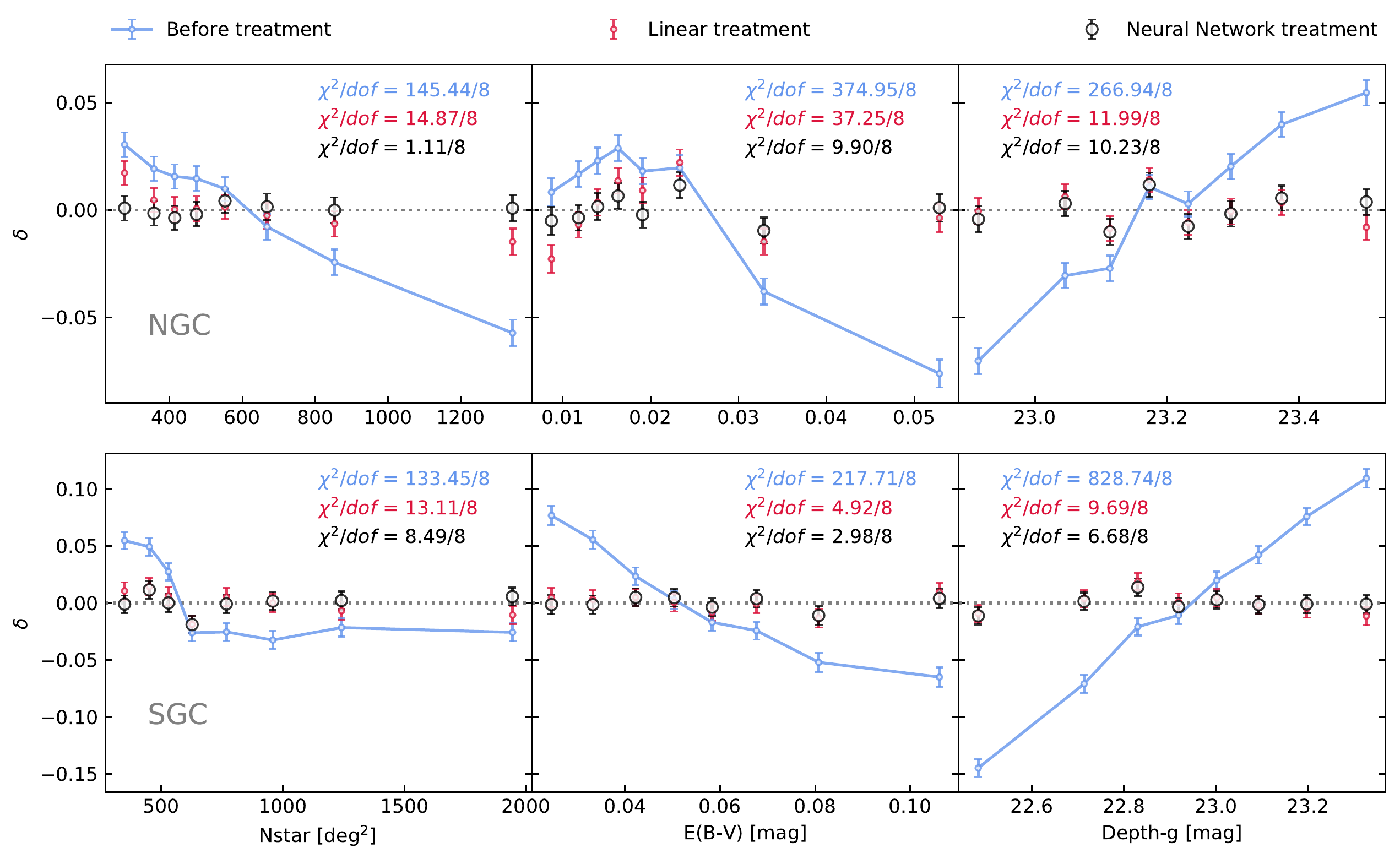}
    \caption{Mean density of the DR16 quasars vs imaging properties before imaging systematics correction (blue), after linear (red), and neural network (black) treatments in the NGC (top) and SGC (bottom) regions. The residual fluctuations are characterized using the $\chi^{2}$ statistics and covariance matrix constructed from the EZmocks.}
    \label{fig:photo_sys}
\end{figure*}

Fig. \ref{fig:photo_sys} illustrates the mean density of quasars versus three primary imaging properties in the NGC (top) and SGC (bottom) regions. The survey $depth-g$ causes the strongest trend in both galactic caps with chi-squared values around $266.94$ and $828.74$, respectively, in the NGC and SGC. The NGC sample also indicates a nonlinear trend against the extinction with $\chi^{2}=374.95$, which remains significant ($\chi^{2}=36.25$) even after the linear treatment (red). On the other hand, our neural network approach (black) is able to reduce the chi-squared error to $9.9$ (see \citealt{RezaiCompanion} for more details). 

%%%%%%%%%%%%%%%%%%%%%%%%%%%%%%%%%%%%%%%%%%%%%%%%%%
%%%%%%%%%%%%%%%%%%% Mocks %%%%%%%%%%%%%%%%%%%%%%%%
%%%%%%%%%%%%%%%%%%%%%%%%%%%%%%%%%%%%%%%%%%%%%%%%%%
\section{Mocks}
\label{sec:mocks}

To estimate the covariance matrix we use a set of 1000 mock catalogues described in \cite{Zhao_2021}, called EZmocks.  These synthetic clustering catalogues include the survey geometry, redshift evolution and systematics, and are based on the framework of \cite{Chuang_2014} using the Effective Zel’dovich approximation. The simulation assumes a $\Lambda$CDM cosmology of $h=0.6777$, $\Omega_m=0.307115$, $\Omega_\Lambda=0.692885$, $\Omega_b=0.048206$, $\sigma_8=0.8225$ and $n_s=0.9611$ in a cube of 5 $h^{-1}$Gpc comoving sidelength. The mocks assume zero \fNL. 

The EZmocks were made from seven simulation snapshots with the clustering signal tuned to match the DR16 quasar catalogue. The EZmocks are fast, approximate mocks that agree well with N-body simulations at large scales, but show some discrepancy at small scales due to non-linear gravitational interactions. These effects only impact scales above $k>0.25\, h/{\rm Mpc}$ which are not used in this study since we only consider only linear scales. Thus we expect no problems for our analysis resulting from the approximations made in creating these simulations.

To  quantify  the  impact  of  systematic  effects  present  in  the  data, we introduced linear fluctuations in the distribution of tracers based on the survey depth in the $g$ band, and the SDSS stellar density map. An angular map was generated based on these two maps and the EZmocks were down-sampled randomly by discarding mock tracers with a probability selected according to this map. Thus, the EZmocks are artificially contaminated with linear systematic trends. Additionally, the effects of redshift failure and fiber collisions were introduced into the mocks (see \citealt{Zhao_2021} for details), however these effects mainly play a role at much smaller scales then used in our analysis. Thus, the contaminated mocks include the uncertainty due to the stochasticity of the systematic treatment. We then assigned weights to the contaminated mocks to account for these effects using the two techniques described Section~\ref{sec:sys}, but using only the two templates used to assign systematics to the mocks. Thus, we can test the effectiveness of the neural network based approaches in comparison to the linear regression method, and the effect of the different approaches, although because they differ in details, we cannot use the mocks to measure an absolute size of the effect of contaminants or their removal.

%%%%%%%%%%%%%%%%%%%%%%%%%%%%%%%%%%%%%%%%%%%%%%%%%%
%%%%%%%%%%%%%%%%%%% Analysis Technique %%%%%%%%%%%
%%%%%%%%%%%%%%%%%%%%%%%%%%%%%%%%%%%%%%%%%%%%%%%%%%
\section{Analysis Technique}  \label{sec:method}

In this section we outline the analysis technique used in this study. The methods are similar to previous clustering studies in Fourier space, however the treatment of the window function and integral constraint differ slightly to ensure a higher precision at very large scales. Additionally, we apply redshift weights designed to optimise constraints on primordial non-Gaussianity that are not used in the standard BAO and RSD studies. We restrict our analysis to linear scales and can therefore model the power spectrum without non-linear corrections.

\subsection{Power spectrum Estimation}
\label{sec:estimator}
We estimate the power spectrum following the approach of \cite{feldman_power-spectrum_1994}, calculating the observed over-density field using a random catalogue with the same survey geometry, completeness and redshift distribution as our data to approximate the expected mean density of the quasar sample. We start with the un-normalised over-density field $F(x)$,
\begin{equation}
F(\mbf{x}) = n_q (\mbf{x}) w_q(\mbf{x}) - \alpha n_r(\mbf{x}) w_r(\mbf{x})
\end{equation}
where $n_q(\mbf{x})$ is the quasar density on a 3D cartesian grid $x$,  $n_r(\mbf{x})$ the density of the random catalog, and $\alpha=\sum w_\mrm{d,noFKP} / \sum w_\mrm{r,noFKP}$ matches weighted densities in the data and random catalogues and $w_q$ and $w_r$ are the weights that correct for various observing effects as discussed in Section~\ref{sec:cat}. The objects in the catalogues are put on a grid of size $L = 6600\, {\rm Mpc}/h$ and $N=512$ using the Cloud-in-Cell interpolation scheme (see \citealt{Sefusattietal:2016} for instance).
We then determine the multipoles of the power spectrum by taking a series of Fast Fourier Transform (FFTs) of $F(\mbf{x})$ and applying the estimator of \cite{Bianchietal:2015} and \cite{Soccimarro:2015}. The monopole, for instance, is then calculated as
\begin{equation}
P_0(k)=\frac{1}{A}\int \frac{d\Omega_k}{4\pi}\left[\hat{F}(\mbf{k})\hat{F}^*(\mbf{k})-S\right]
\end{equation}
where $A$ the normalisation and the short noise $S$ are given by
\begin{eqnarray}
 A &=& \alpha \sum w^2_\mrm{r,FKP} \ n(z) \ w_\mrm{r,noFKP}   \\
 S &=& \sum w^2_\mrm{q} + \alpha^2 \sum w^2_\mrm{r}
\end{eqnarray}
where the subscripts ``q'' and ``r'' refer to quasars and randoms, respectively. Note that the values of $\alpha$, the normalisation $A$ and shot noise $S$ differ slightly from previous analyses due to the treatment of the completeness in the random catalogue since the randoms are weighted instead of down-sampled to match the data catalogue. We cut off our analysis at $k = 0.12\, h/{\rm Mpc}$, well below the Nyquist frequency of $k_\mrm{Ny}=0.24\, h/{\rm Mpc}$ and bin the power spectrum using $\Delta k = 2 k_f = 4\pi/L$. This procedure is the same as in the BAO and RSD studies in Fourier space of the quasar \cite{Neveux-eboss-qso-pk}, LRG \cite{gil-marin20a} and ELG \cite{demattia20a} samples.

\begin{figure*}
    \centering
    \includegraphics[width=0.45\textwidth]{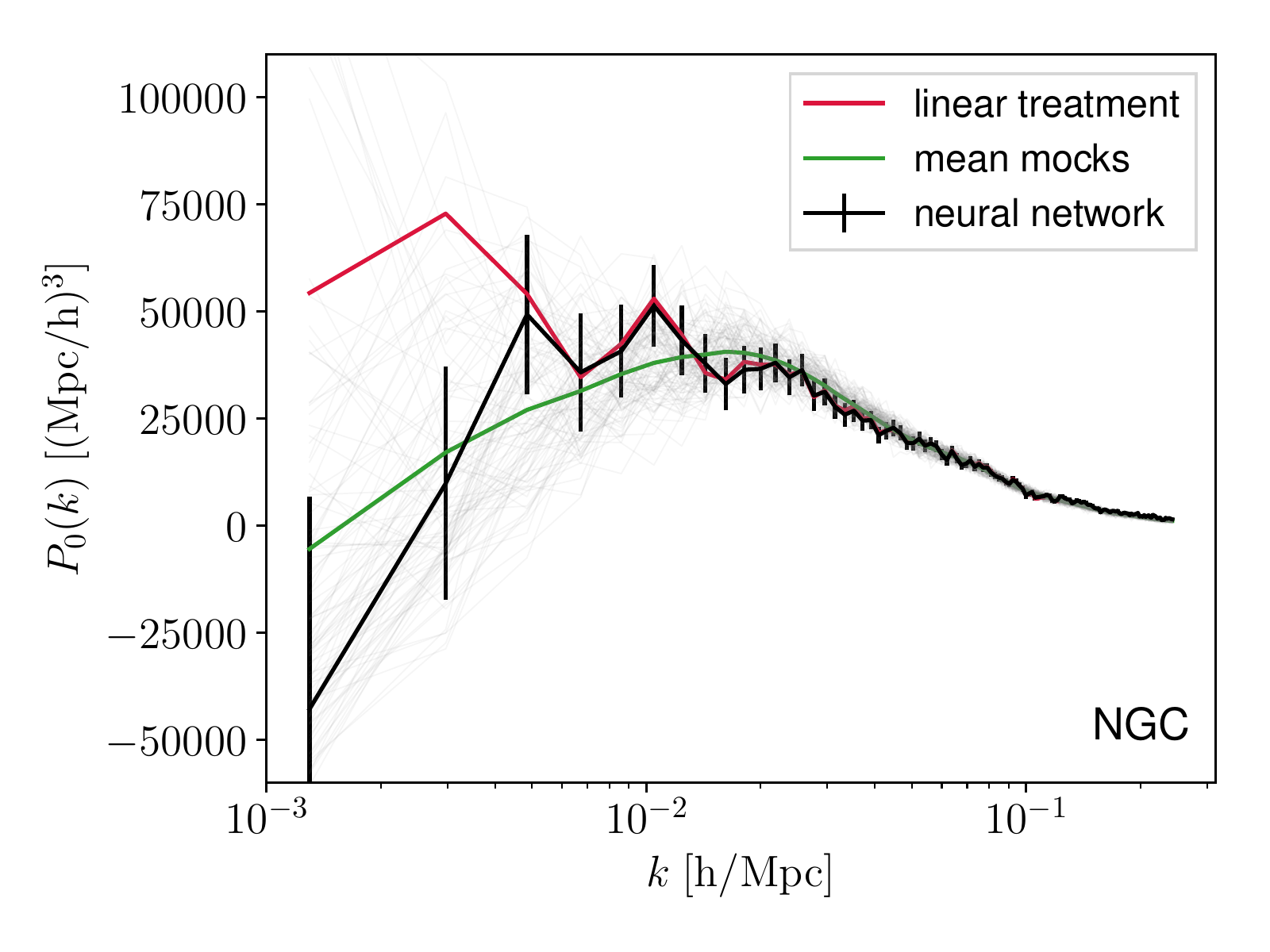}
    \includegraphics[width=0.45\textwidth]{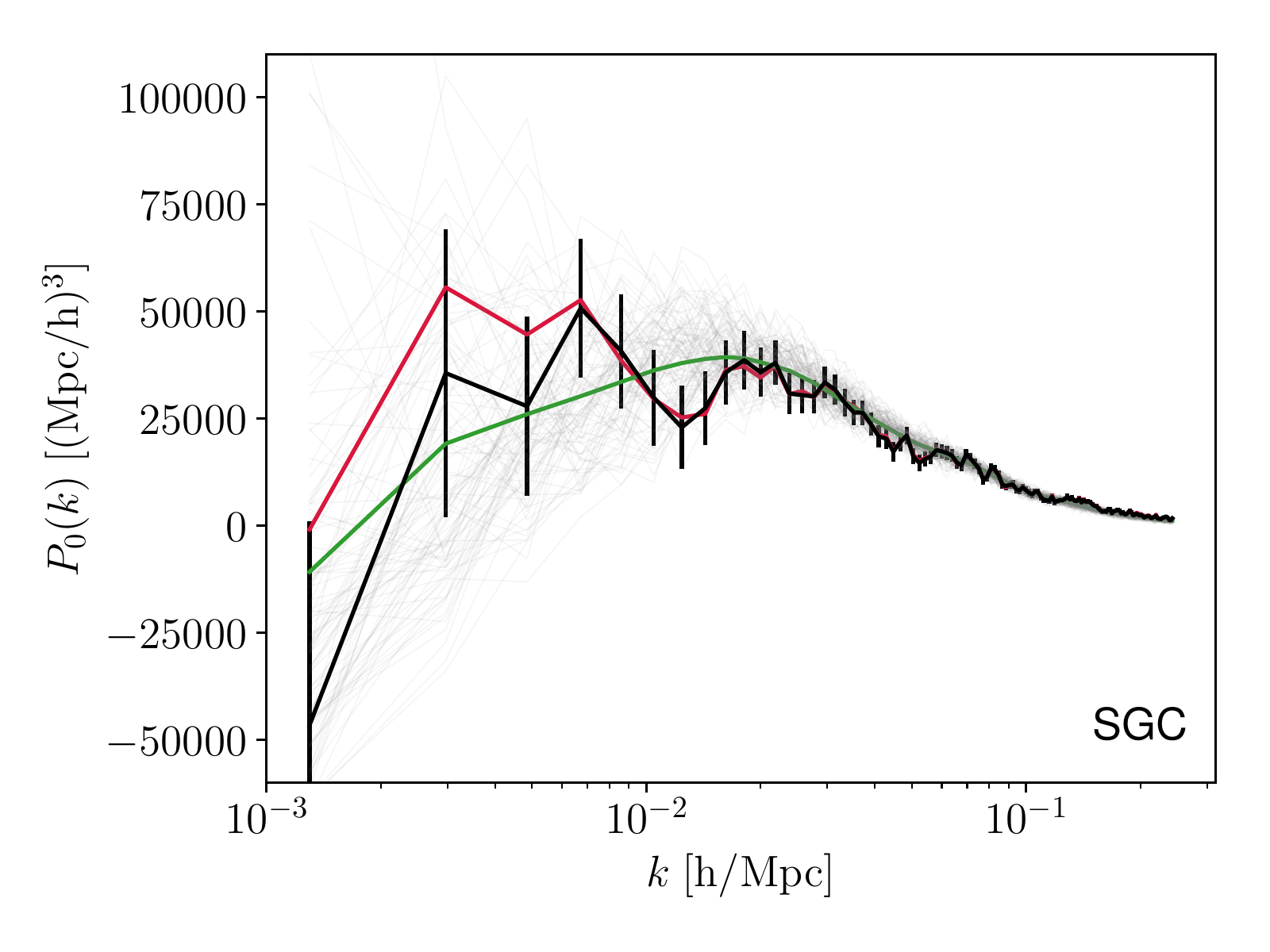}
    \caption{Power spectrum monopole of the eBOSS DR16 QSO NGC (left) and SGC (right) for different systematic treatments. The red line corresponds to the standard linear treatment for systematics applied to the data, while the black line represents the advanced neural network based systematic removal technique. The mean of the mocks is displayed in green, while the individual mocks are over-plotted in grey.}
    \label{fig:Pk}
\end{figure*}

Figure~\ref{fig:Pk} shows the power spectrum monopole for the NGC (left) and SGC (right) regions after the neural network (black) and linear treatment (red). To visualise the variance and expected mean at large scales, the power spectrum result for the individual mocks is over-plotted in grey and the mean of the mocks in green. We do not use the quadrupole in this analysis since it is very noise at large scales for the DR16 QSO data.

\subsection{Window function and Integral constraint}

The area of the survey leaves a footprint in the clustering of quasars and needs to be taking into account when analysing two points statistics. In Fourier space, the estimated power spectrum from the survey is not just the quasar power spectrum but the power spectrum convolved with the survey window function. We include this effect by convolving the model power spectrum with the window function. We calculate the window function multipoles using random pair-counts as outlined in \cite{beutler_clustering_2017} and \cite{wilson_good_2016} with the normalisation of the window given by the same normalisation as the power spectrum. 
%Figure \ref{fig:window} shows the first three window function multipoles in configuration space used in this analysis.
The convolved power spectrum $\hat{P}(k)$ is then calculated from the convolved correlation function $\hat{\xi}$ through
\begin{equation}
    \hat{P}_l= 4 \pi (-i)^l \int ds \ s^2 \hat{\xi}_l(s) j_l(sk).
\end{equation}

Additionally, we have to account for the integral constraint. The formalism described in Section~\ref{sec:estimator} to estimate the over-density field $F(\mbf{x})$ implicitly sets the mean quasar density to zero in our analysis. Taking a Fourier transform of $F(\mbf{x})$ then gives rise to an additional contribution to the monopole through the convolution with the window function
\begin{equation}
    \left< F(\mbf{k})F^*(\mbf{k}) \right> = \int \frac{dk'}{(2\pi)^3} \left[ \tilde{P}(k') - \tilde{P}(0) \delta_D(k) \right] W(k) 
\end{equation}
with the shot-noise uncorrected power spectrum $\tilde{P}$. As pointed out by \cite{de_Mattia_2019}, there is an additional contribution to the integral constraint because the radial distribution of the randoms is exactly matched to the data giving rise to an `radial' contribution to the integral constraint. Here we account for the radial integral constraint by calculating the mean power spectrum of the mocks with the redshifts of the randoms drawn for the data mocks, i.e. with the radial integral constraint introduced into the mocks, and with the redshifts drawn from the $n(z)$-distribution. The additional radial component is then calculated as the fractional difference of the mean power spectra, $W_\mathrm{RIC}$. The total integral constraint is then given by $- \tilde{P}(0) W(k) -\tilde{P}(k) W_\mathrm{RIC}$.    

\subsection{Optimal Weights}  \label{sec:OptimalWeights}
To fully exploit all the information in the LSS of the quasar clustering we apply redshift weights designed to minimise the uncertainty in the PNG parameter \fNL as derived in \cite{mueller_clustering_2018}. The redshift weighting technique was first developed for BAO measurements \citep{zhu_optimal_2015,zhu_redshift_2016} and was subsequently also applied to optimised RSD measurements \citep{ruggeri_optimal_2017,Zhao-rsd-optimal-2019}. While the FKP weights balance shot noise and cosmic variance contributions to minimise the statisical uncertanity, the redshift weights take the redshift evolution of the underlying physical theory into consideration to further improve the constraints. This methods avoids a binning in redshift space and reduces the information loss at the edges of the bins. In this study we only focus on the power spectrum monopole and thus the redshift weight is defined as
\begin{equation}
    w_z^2(z) = (b(z)+\frac{1}{3}f(z))(b-p)D(z). 
\end{equation}
with the fiducial bias given by $b(z)=0.53 + 0.29(1 + z)^2$. We apply these weights to each object in the data and randoms. The total weight is then
\begin{equation}
    w_\mathrm{tot}=w_\mathrm{FKP} \times w_\mrm{CP} \times w_\mrm{NOZ} \times w_z \times w_\mrm{obs}.
\end{equation}
The effective redshift of $z_\mrm{eff}= (\sum  w_\mathrm{tot} z)/ \sum w_\mathrm{tot}=1.51$ increases to $z_\mrm{eff}=1.80$, reflecting the up-weighting to higher redshift due to the redshift weights.

\subsection{Power spectrum Model}
\begin{figure}
    \centering
    \includegraphics[width=0.45\textwidth]{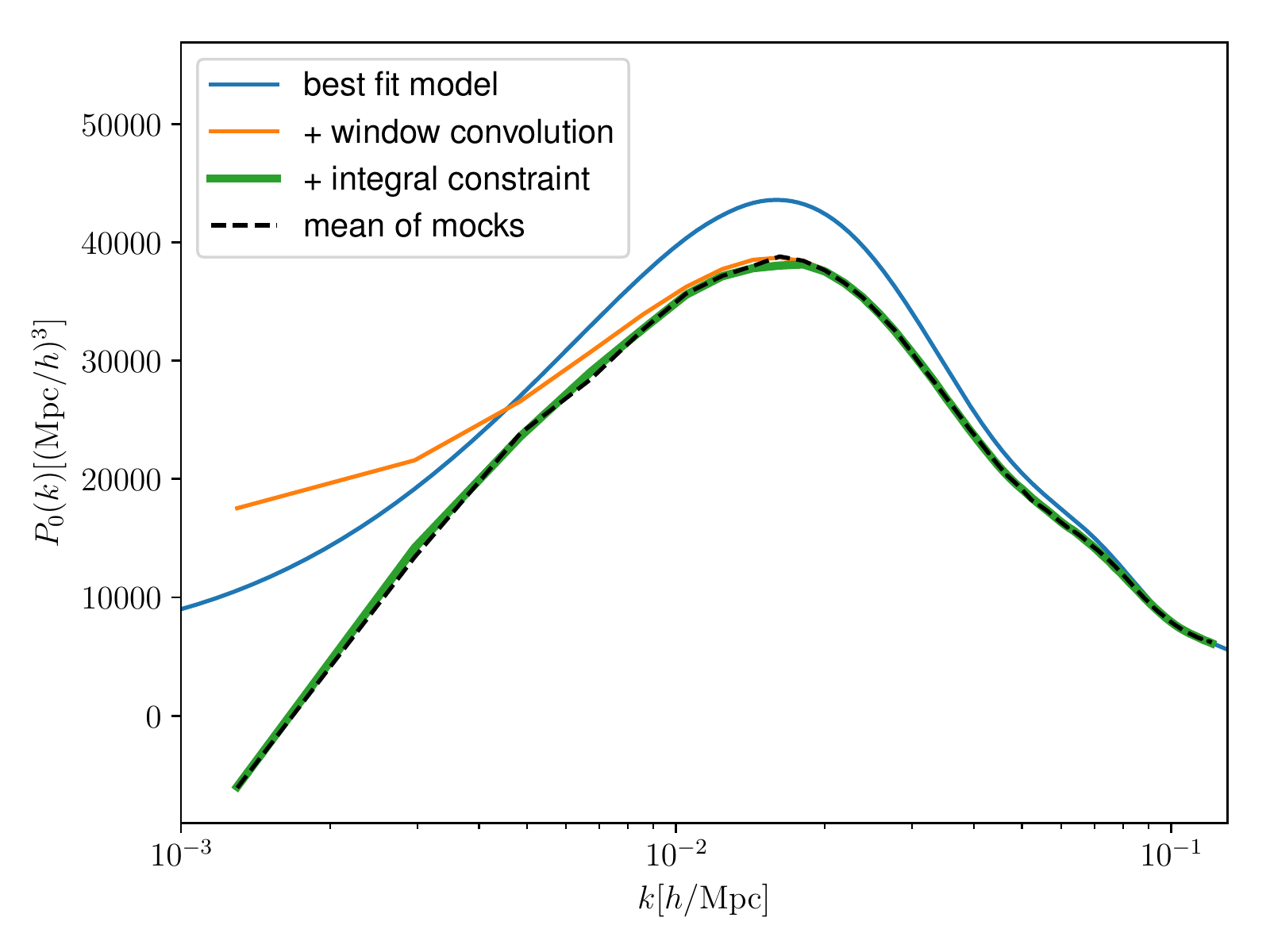}
    \caption{Theoretical model of the power spectrum monopole given the best fit parameters to the mean of the mocks (blue), with the convolution of the window function (orange) and with the integral constraints (green) applied to the model. The black dashed line represents the mean of the mocks. }
    \label{fig:model}
\end{figure}

We restrict our analysis to linear scales with the redshift power spectrum, $P(k,\mu)$ modelled as
\begin{equation}
    P(k,\mu)=\frac{(b+\mu^2 f)^2P_m(k)}{1+(k\mu \sigma_v/H_0)^2}
\end{equation}
with the linear matter power spectrum $P_m(k)$ from CAMB, the galaxy bias $b$, the growth rate $f$ and a Lorenzian damping factor given the velocity dispersion $\sigma_v$.
The power spectrum multipoles can then be calculated as
\begin{equation}
    P_\ell(k)=\frac{2\ell+1}{2}\int_{-1}^{1}P(k,\mu)L_\ell(\mu)d\mu .
\end{equation}
Figure~\ref{fig:model} shows the model of the power spectrum monopole (blue) using the best fit parameters from the fit to the mean of the mocks. When the convolution with the survey window function is included (orange) the power spectrum is modified at large scales. If we additionally apply the integral constraint to the model (green), we can recover the mean of the mocks (black dashed).

\subsection{Covariance and Likelihood}  \label{sec:cov}
For the main quasar clustering catalogue we estimate the covariance of the power spectrum multipoles numerically from the 1000 EZ-mocks described in Section~\ref{sec:mocks}. 
We invert the covariance including the Hartlap correction \citep{Hartlap07}, a 6.5\% effect for our configuration. The constraints on $f_\mrm{NL}$ along with the nuisance parameters $b$ and $\sigma_v$ are then inferred from maximizing the likelihood $\mathcal{L} \propto e^{-\chi^2/2}$ 
using a Monte Carlo Markov Chain (MCMC) Metropolis-Hasting algorithm with the $\chi^2$ given by $\chi^2=( \mbf{D} - \mbf{T(p)} ) ^T \mbf{C}^{-1}( \mbf{D} - \mbf{T(p)})$ where $D$ refers to the data vector, $T(p)$ to the theory prediction and p the set of parameters. We use flat priors on all parameters and to assure convergence of the MCMC chains we apply the Gelma-Rubin statistics. All MCMC chains have a converge of $R<0.01$ or lower. 
In our analysis we marginalised over the shot noise S, the bias and the velocity dispersion $\sigma_v$ in each cap separately as nuisance parameters. With the rest of the cosmological parameters fixed, we vary four parameters in total in our MCMC runs.

\section{Robustness tests with mocks}  \label{sec:tests}
\subsection{Testing the model}

\begin{figure}
    \centering
    \includegraphics[width=0.45\textwidth]{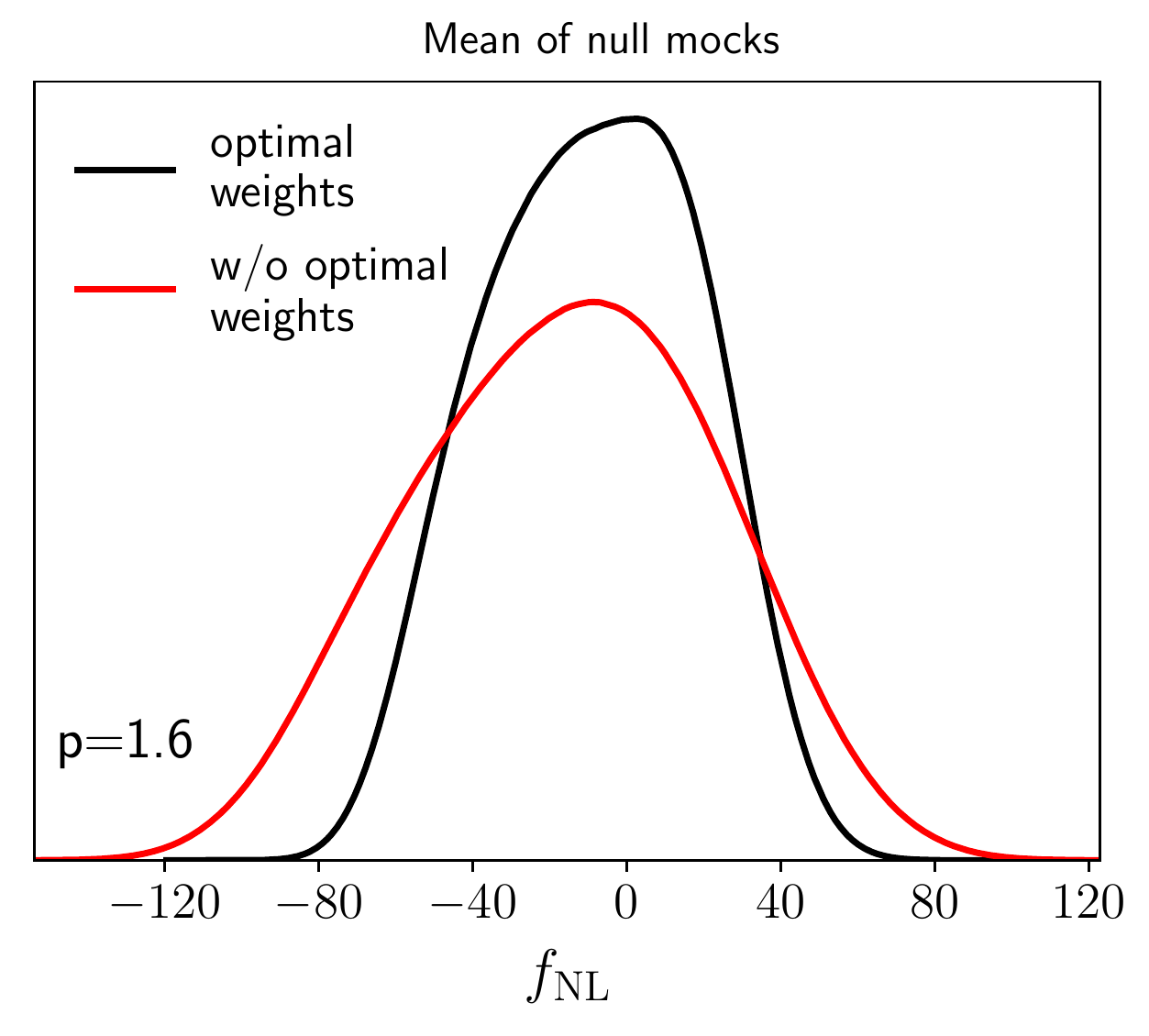}
    \caption{Marginalized posterior distribution of \fNL from the mean power-spectrum monopole of the null mocks but assuming the data covariance. Results for the null mocks with (black) and without (red) optimal weights.}
    \label{fig:fNLmocksmean}
\end{figure}

We test the reliability of our analysis pipeline, in particular, the modelling of the window function and integral constraints, using the EZ-mocks described in Section~\ref{sec:mocks}, with no observational systematics added, which we call `null' mocks.
\begin{figure*}
    \centering
    \includegraphics[width=1.0\textwidth]{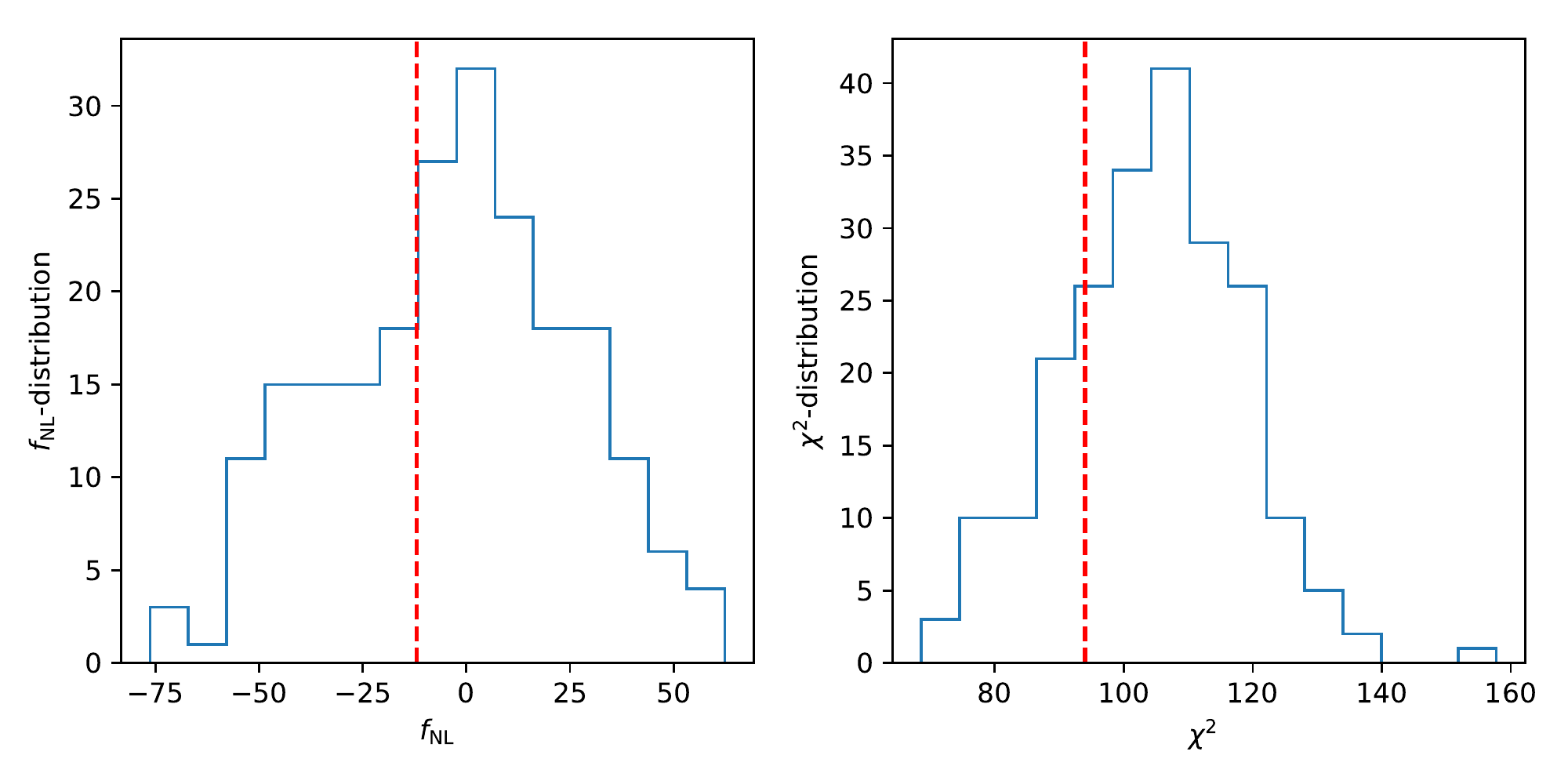}
    \caption{Robustness test I: Best-fit \fNL (left) and corresponding $\chi^2$ (right) from performing the MCMC analysis for 200 null mocks constructed without systematic errors and including optimal weights in the analysis. The red dashed lines correspond to the best fit results from the data.}
    \label{fig:fNLmocks}
\end{figure*}
Figure~\ref{fig:fNLmocks} shows the minimum $\chi^2$-distribution (right) and the best-fit \fNL (left) of 200 individual mocks. We recover zero \fNL as expected from the mocks and find no bias within our pipeline. The red dashed line shows the best fit $\chi^2$ and the best fit \fNL value of the data. 
 Figure~\ref{fig:fNLmocksmean} shows the constraints on $f_\mrm{NL}$ from the mean power spectrum of the 1000 null mocks with the optimal weights (black) and without (red). We can recover zero non-Gaussianity within 1-$\sigma$ and find no significant bias in the pipeline. 

%\begin{figure}
%    \ovarainceentering
%    \includegraphics[width=0.45\textwidth]{plots/fNL_mocksmean_v7.png}
%    \caption{Robusteness test II: Best-fit \fNL on the mean of 1000 mocks for the NGC (black), %SGC (red) and combined NGC+SGC (blue).}
%    \label{fig:fNLmocksmean}
%\end{figure}

\subsection{Observational systematics}

\begin{figure*}
    \centering
    \includegraphics[width=0.45\textwidth]{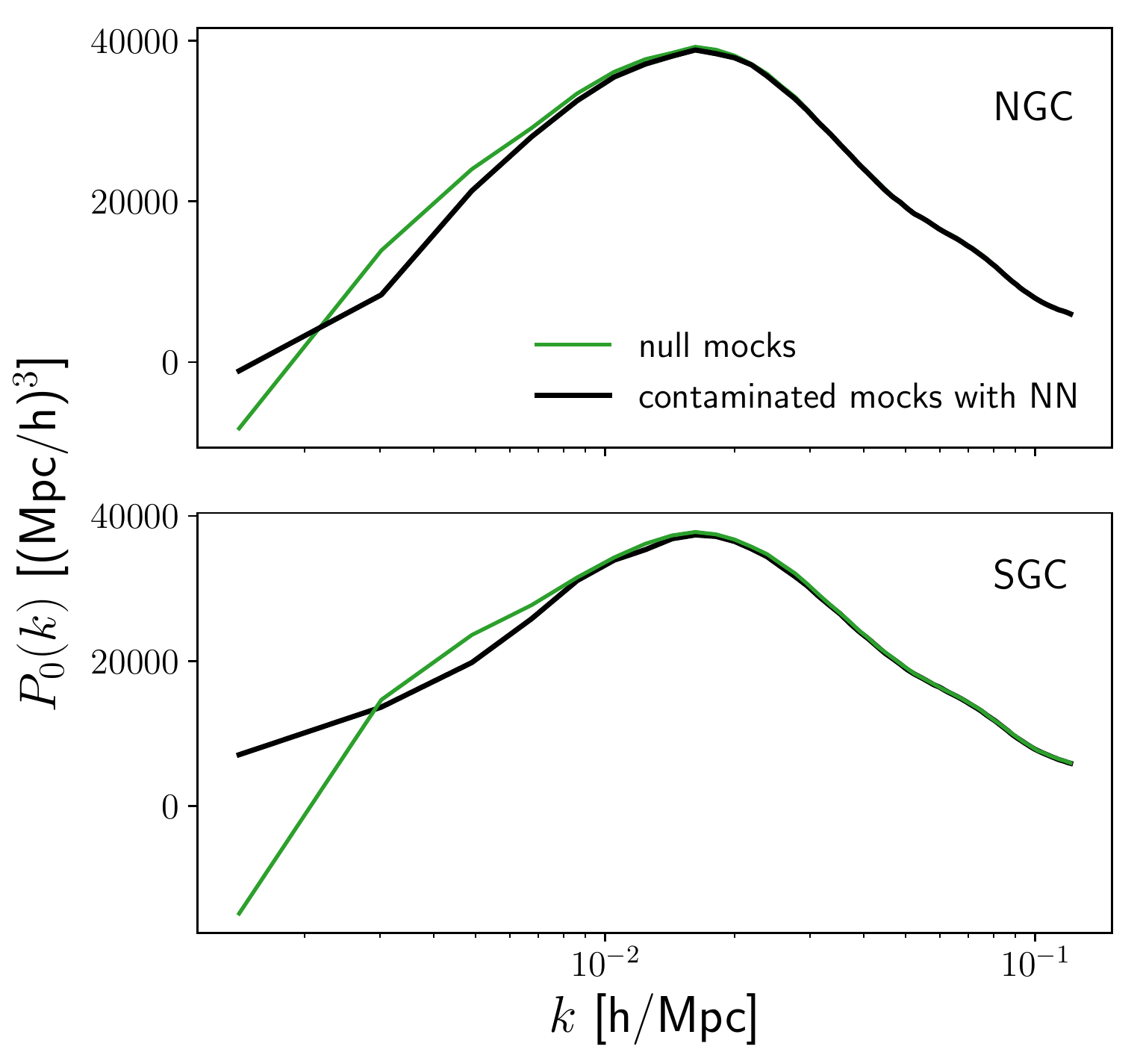}
    \includegraphics[width=0.45\textwidth]{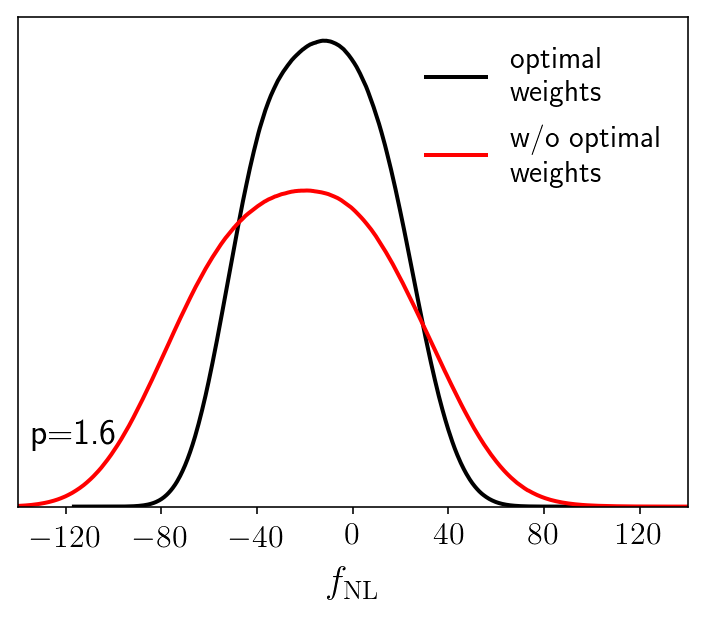}
    \caption{Impact of the treatment of systematics on the mocks: Left panel: Mean power spectrum monopole from the null mocks (green) and contaminated mocks (black) with the neural network removal technique applied to each individual mock. Right panel: Corresponding \fNL constraint from the mean of the contaminated mocks (with errors as for a single mock).}
    \label{fig:cont_mocks}
\end{figure*}

Using mocks with and without linear systematics added, as described in Section~\ref{sec:mocks}, we now test our ability to recover the correct value of \fNL. The biggest concern is that, unfortunately any treatment of systematic to remove contamination also has the potential to remove part of the signal and decrease the power especially at large scales, where there are only very few modes. Here we study this effect by applying the neural network based correction to the contaminated mocks and compare the results to the null mocks. The left-hand panel of Figure~\ref{fig:cont_mocks} shows the mean power spectrum monopole of the contaminated mocks after neural net removal of systematics (black) and the null mocks (green) for the NGC (upper panel) and SGC (lower panel). The power spectrum at large scales has a reduced power after the neural network treatment compared to the null, i.e. uncontaminated, mocks. This consequentially shifts the \fNL maximum likelihood to negative values resulting in a mild bias of the \fNL constraints. Note, that we do not use the first $k$-bin in the MCMC analysis due to its instability to the systematic correction. The maximum likelihood value is $f_\mathrm{NL}=-12$ with optimal weights and $f_\mathrm{NL}=-19$. This gives us an estimate of the overall bias caused by the systematic treatment.

\citet{RezaiCompanion} present a method to correct for this over-subtraction, calibrated on the mocks. This removes this small bias. However, the systematic errors are only applied to the mocks using a linear model with two templates, and the effect of the NN correction is significantly smaller than for the data. Thus, we do not trust this correction when applied to the data. Given the small amplitude of this effect, we choose to not apply a correction, rather than apply an incorrect correction.

%\subsection{Skewness}
%The skewness $G(k)$, defined as
%\begin{equation}
%G(k) = \frac{\sqrt{N(N-1)}}{N(N-2)} \sum_{n=1}^{N} \left( \frac{X(k) %- \bar{X}(k)}{\sigma} \right)^3,
%\end{equation}
%describes the asymmetry of the probability distribution of a random variable $X(k)$ around its mean $\bar{X}(k)$ with standard deviation $\sigma$ (given by $\sigma^2 = \frac{6N(N-1)}{(N-2)(N+1)(N+3)}$ for a Gaussian dsitribution). Following the analysis of cite{Ashley}, Figure X shows the skewness for $X(k)=\{P(k), \mrm{ln}(Pk)\}$ using the 1000 EZ mocks. 

\section{Results}  \label{sec:results}
In this section we present the results of measuring the local, primordial non-Gaussianity parameter \fNL, summarised in Table~\ref{tab:Results}. We discuss the impact of optimal weights to improve the constraints, the impact of photometric systematics and the merging history of halos. Next, we evaluate the robustness of our results by varying our default analysis assumptions including the theoretical model as well as scale cuts and priors (see Table~\ref{tab:Robustness}).

\subsection{Constraints on the local PNG}

\begin{figure}
 \centering
    \includegraphics[width=0.45\textwidth]{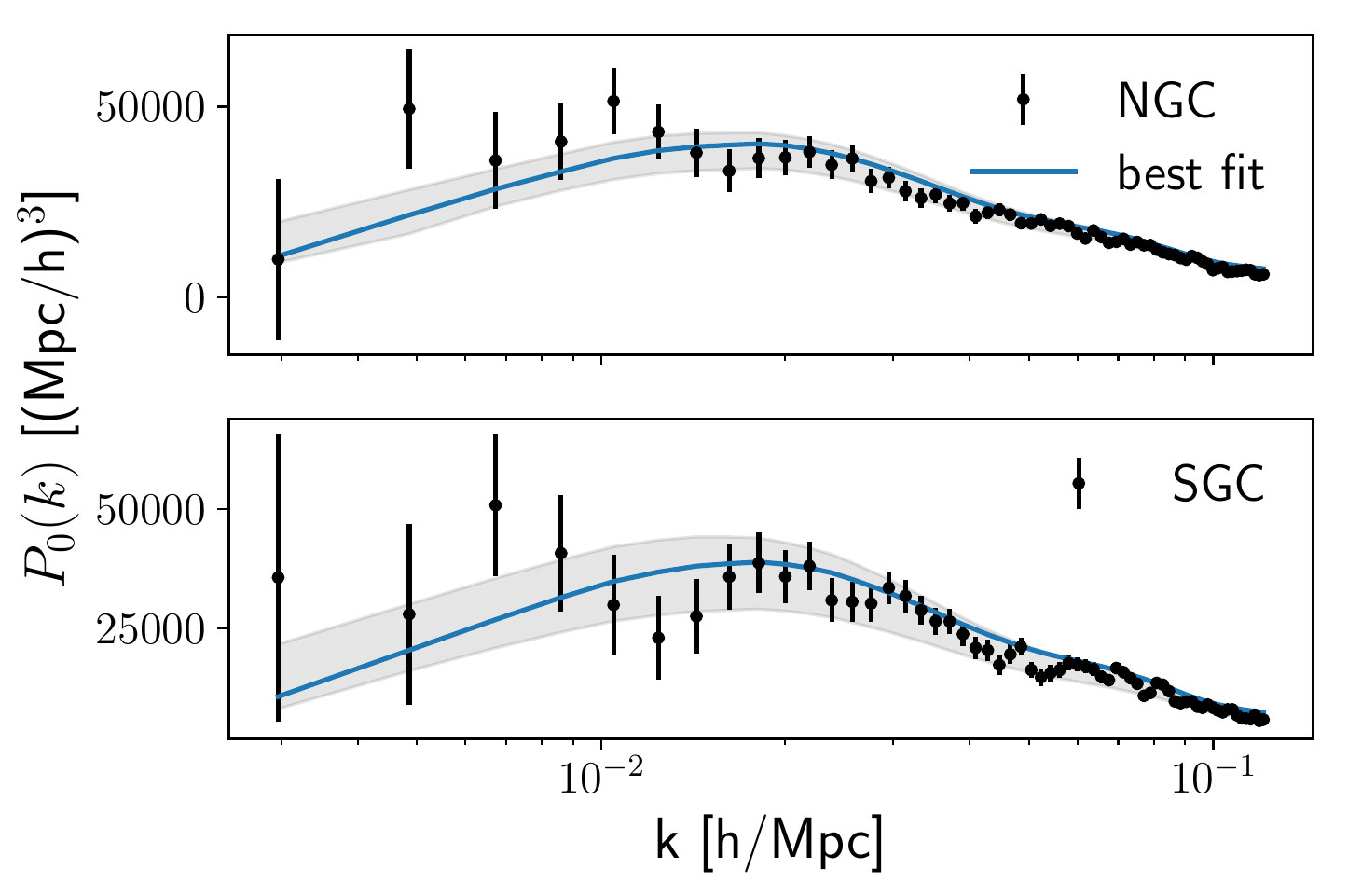}
    \caption{Power spectrum monopole for the NGC [upper panel] and SGC [lower panel]. The best fit model is plotted in blue. The blue contours show the range of models within the 1-$\sigma$ uncertainty in \fNL.}
    \label{fig:BestFit}
\end{figure}

We constrain the local PNG parameter \fNL to $-33<f_\mathrm{NL}<10$ at 68\% C.L and $-51<f_\mathrm{NL}<28$ at 95\% C.L using the eBOSS quasar sample with a redshift range of $0.8<z<2.2$ and a minimum scale of $k_\mathrm{min}=0.019\, h/{\rm Mpc}$ included in the analysis. These are the tightest constraints on PNG from a spectroscopic survey to date covering the largest survey volume and using information on the largest cosmological scales. We find no evidence of PNG with the \fNL parameter agreeing with zero within the 1-$\sigma$ uncertainty.  We improve upon the eBOSS DR14 results of $-81 < f_\mathrm{NL} < 26$ at 95\% C.L by 35\%. 
The default settings for our analysis are summarised in Table~\ref{tab:defaults}. Figure~\ref{fig:BestFit} shows the power spectrum monopole in the NGC (upper panel) and SGC (lower panel) with the best fit model (blue line) and the 1-$\sigma$ uncertainty in \fNL (grey contours). Overall, the best fit of our model to the data is very good with $\chi^2$/dof = 94/119.

\begin{figure*}
 \centering
 \begin{subfigure}[t]{0.5\textwidth}
    \centering
    \includegraphics[height=3.3in]{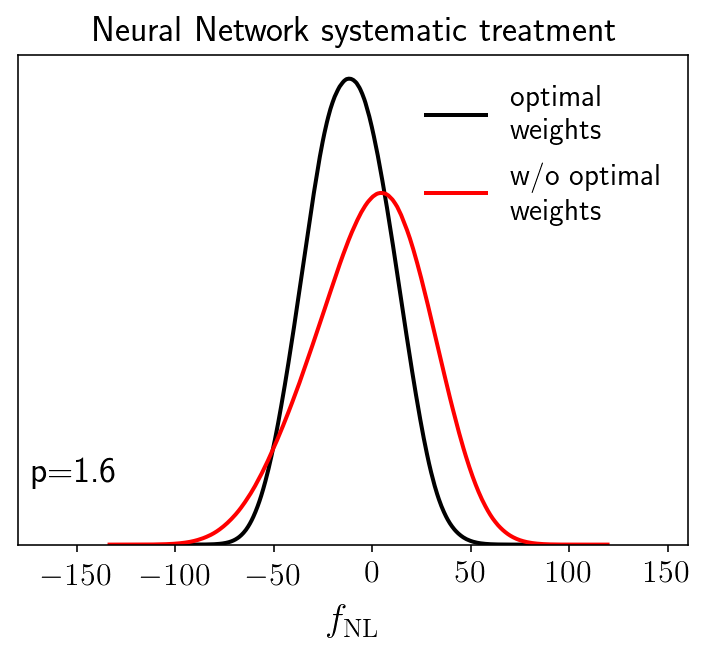}
    %\caption{'Fair' with Optimal Weights}
 \end{subfigure}%
 \begin{subfigure}[t]{0.5\textwidth}
  \centering
    \includegraphics[height=3.3in]{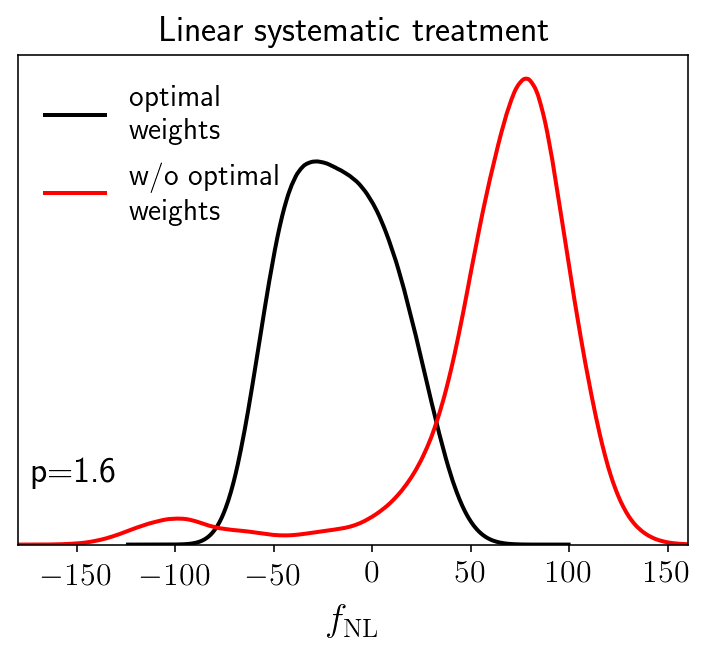}
    %\caption{'Naive' with Optimal Weights}
 \end{subfigure}
  \begin{subfigure}[t]{0.5\textwidth}
   \centering
    \includegraphics[height=3.3in]{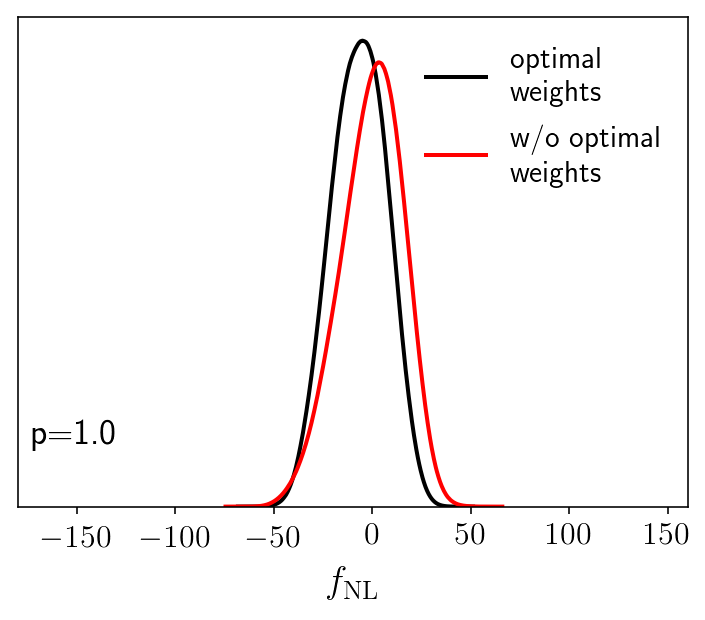}
    %\caption{'Fair' with Optimal Weights}
 \end{subfigure}%
 \begin{subfigure}[t]{0.5\textwidth}
  \centering
    \includegraphics[height=3.3in]{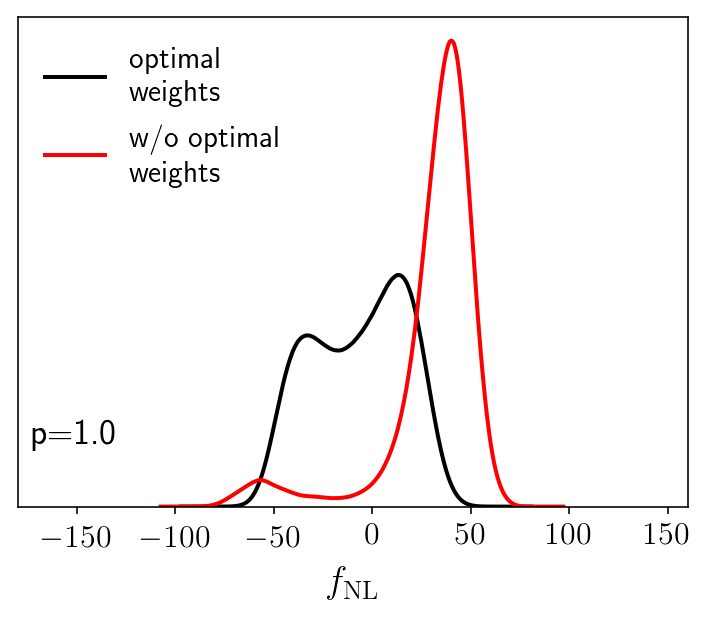}
 \end{subfigure}
 \caption{Marginalized posterior distribution of \fNL for the Neural Network systematic treatment [left] and linear systematic treatment [right]. Lower panels for p=1 assuming no recent merger history and upper panels for p=1.6. These are our main results from the DR16 QSO data.}
\label{fig:results}
 \end{figure*}

Here we have applied a set of optimal weights including redshift weights and optimal FKP weights as discussed in Section~\ref{sec:OptimalWeights}. The redshift weighting technique includes information of the underlying theoretical model and increases the effective redshift of the sample from $z_\mathrm{eff}=1.51$ to $z_\mathrm{eff}=1.82$. Since the impact of PNG is stronger at high redshift the constraints can be improved. Additionally the FKP weight is chosen at $P_0 = 20 000 \ (\mathrm{Mpc}/h)^3$ instead of the standard $P_0 = 60 00 \ (\mathrm{Mpc}/h)^3$ that is used in BAO and RSD analysis to balance shot noise and cosmic variance at large scales, which are of interest in this study. 
Figure ~\ref{fig:results} shows the the constraints on \fNL with (black line) and without (red line) the optimal weights; our results using the NN weights and merging factor $p=1.6$ are shown in the upper-left panel. The optimal weights lead to an improvement of 37\% compared to the constraint without optimal weights of $-28<f_\mathrm{NL}<31$ at 68\% C.L and a mild shift to higher \fNL values. The redshift weighting technique is particularly effective for the quasar sample because of its large redshift range but also because of the strong redshift dependency of the quasar bias.  

Next, we study the effect of the photometric contaminations in the sample and different systematics removal techniques. The photometric systematics in particular effect the largest scales of the quasar power spectrum monopole (see Figure~\ref{fig:Pk}) and their correct removal is crucial for any study of PNG. 
 In this study we focus on two systematic treatments:
\begin{itemize}
\item Linear treatment: standard linear weights as used in the BAO analysis
\item Neural network treatment: advanced machine learning based approach
\end{itemize}
In Section~\ref{sec:sys} we summarise each technique for the data. The details of the systematic treatment used for this analysis discussed in detail in the accompanying paper of \citet{RezaiCompanion}. 
The main differences between the two approaches is that the neural network technique allows us to remove non-linear effects caused by the photometric contaminations while the standard weights only remove linear trends. Additionally, \citet{RezaiCompanion} identified that the stellar density was better matched using a new template constructed using Gaia data, whereas the standard treatment used a template based on SDSS data (see Section~\ref{sec:NN_weights}).  Figure~\ref{fig:results} shows the constraints on \fNL for the linear systematics treatment. With the optimal weights, the linear \fNL measurement is in agreement with zero at 1-$\sigma$ and is consistent with the neural network systematics removal. However, without the optimal weights the maximum likelihood value of \fNL shifts by almost 4 $\sigma$, hinting at systematic contamination at low redshifts. For the linear systematic removal, our constraints on \fNL are not robust and confirm the findings of residual systematics in the standard catalogue as found in \cite{RezaiCompanion}. This potential bias in \fNL highlights the need for robust systematics removal techniques to account for photometric contaminants. 

\begin{figure}
    \centering
    \includegraphics[width=0.45\textwidth]{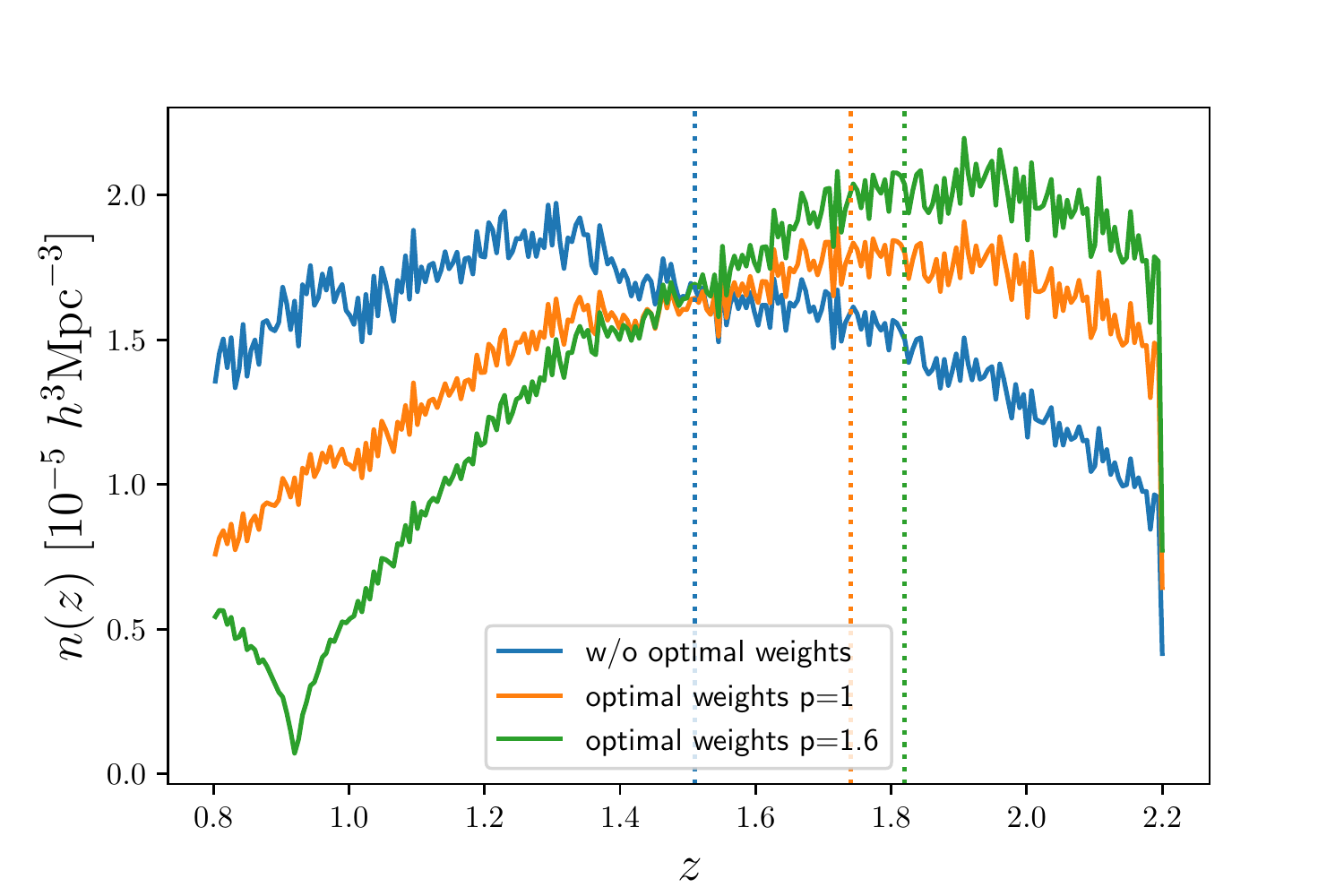}
    \caption{Redshift distribution of the eBOSS QSO sample in the NGC without optimal weights (blue), with optimal weights for $p=1$ (orange) and optimal weights with $p=1.6$ (green). The optimal weights increases the effective redshift from 1.51 to 1.74 ($p=1$) and 1.83 $p=1.6$, shifting the redshift  distribution to peak at higher redshifts. The parameter $p$ depends on the assumed merger history of the QSO sample and defines the scale dependency of the PNG contribution to the halo bias.}
    \label{fig:nz}
\end{figure}

Third, we discuss the impact of the merger history of quasars on the \fNL constraints parametrised as $p$ in Eq.~\ref{eq:bias_ng}. Tracers that mainly populate recently merged halos lead to a weaker scale dependency of the halo bias with $p\sim 1.6$. While this is our default assumption, it is likely that not all quasars are in halos that resulted from recent mergers. Our default results can therefore be considered the most conservative estimate. If we instead assume $p=1$, i.e. no recent merger history, our constraints tighten to $-18<f_\mathrm{NL}<11$ with the optimal weighing technique. This is the most optimistic estimate of the \fNL uncertainties. Note, that the effective redshift in this case is $z_\mathrm{eff}=1.74$ and the improvement of the optimal weights compared to the standard analysis without optimal weights is 10\%. The redshift evolution of the optimal weights with $p=1$ is weaker that for $p=1.6$ resulting in less improvement and a lower overall effective redshift. Figure \ref{fig:nz} shows the redshift distribution, $n(z)$ for different weights. 

Nevertheless, the constraints for $p=1$ are 80\% tighter than for $p=1.6$. Therefore, our 1-$\sigma$ results are between $\pm14<\sigma(f_\mathrm{NL})<\pm21$. Note, that the maximum likelihood value for \fNL does not change with p, only the uncertainty on \fNL varies with p. 

\begin{table}
	 \centering
    	\begin{tabular}{|l|l|}
	\hline
	\hline
	survey & eBOSS \\
	tracer & quasars \\
	redshift range & $0.8<z<2.2$ \\ 
	effective redshift & $z_\mathrm{eff} =1.82$ \\ 
	scale range in $h$/Mpc & $0.0019<k<0.121$ \\
	k bins & 63 with $\Delta k = 2 \pi / L \approx 0.019$ $h$/Mpc \\
	photometric treatment &  neural network weights \\
	nuisance parameters & \{$b_N$, $b_S$, $\Delta P^\mathrm{shot}_N$, $\Delta P^\mathrm{shot}_S$, $\sigma_{v,N}$, $\sigma_{v,S}$\} \\
	merger rate p & 1.6 \\
	Weights & redshift and optimised FKP weights \\
	\hline
	\end{tabular}
	\caption{Default assumptions of the analysis}
	\label{tab:defaults}
\end{table}

\begin{table*}
	 \centering
    	\begin{tabular}{|c|c|c|c|c|c|c|c|}
	\hline
	\hline
	p&Systematic treatment & \fNL 68\% C.L.& \fNL 95\% C.L & Max. Likelihood & Best fit & Marg. Mean & $\chi^2/\mathrm{dof}$\\
	\hline
	\multirow{4}{2em}{1.6}& \textbf{Neural Network}  & $\mathbf{-33<f_\mathrm{NL}<10}$ &$\mathbf{-51<f_\mathrm{NL}<28}$  & \textbf{-12} & \textbf{-13} & \textbf{-12} & \textbf{94/119} \\
	&Linear & $-50<f_\mathrm{NL}<10$ & $-68<f_\mathrm{NL}<36$& -26 & -6 & -17 & 87/119 \\
	\cline{2-8} 
 & w/o Optimal Weights & \\
	 & Neural Network  & $-28<f_\mathrm{NL}<31$ &$-61<f_\mathrm{NL}<51$  & 4 & 11 & -1 & 99/119 \\
	&Linear & $47<f_\mathrm{NL}<103$ & $-114<f_\mathrm{NL}<133$& 78 & 80 & 62 & 89/119 \\
	\hline
	\multirow{4}{2em}{1}&Neural Network & $-17<f_\mathrm{NL}<11$ & $-31<f_\mathrm{NL}<22$& -2 & 1 & -4 & 98/119 \\
	&Linear  & $-39<f_\mathrm{NL}<25$ &$-51<f_\mathrm{NL}<33$  & 13 & 13 & -7 & 87/119 \\
	\cline{2-8}
 & w/o Optimal Weights & \\
	 & Neural Network & $-14<f_\mathrm{NL}<18$ &$-32<f_\mathrm{NL}<29$  & 3 & 10 & 0 & 99/119 \\
	&Linear & $25<f_\mathrm{NL}<52$ & $-65<f_\mathrm{NL}<65$& 40 & 44 & 31 & 89/119 \\
	\hline
	\end{tabular}
	\caption{68\% and 95\% C.L.  constraints on local PNG for the systematic treatment using neural networks as well as the standard removal technique using linear weights. We present the results with the optimal weights and without, as well as for different assumptions on the recent merger history parametrised by p. These results are also displayed in Figure~\ref{fig:results}. The first row, showing the Neural Network treatment with optimal weights is our default result.}
	\label{tab:Results}
\end{table*}

\subsection{Robustness of the results}
\begin{figure*}
 \centering
    \includegraphics[width=0.9\textwidth]{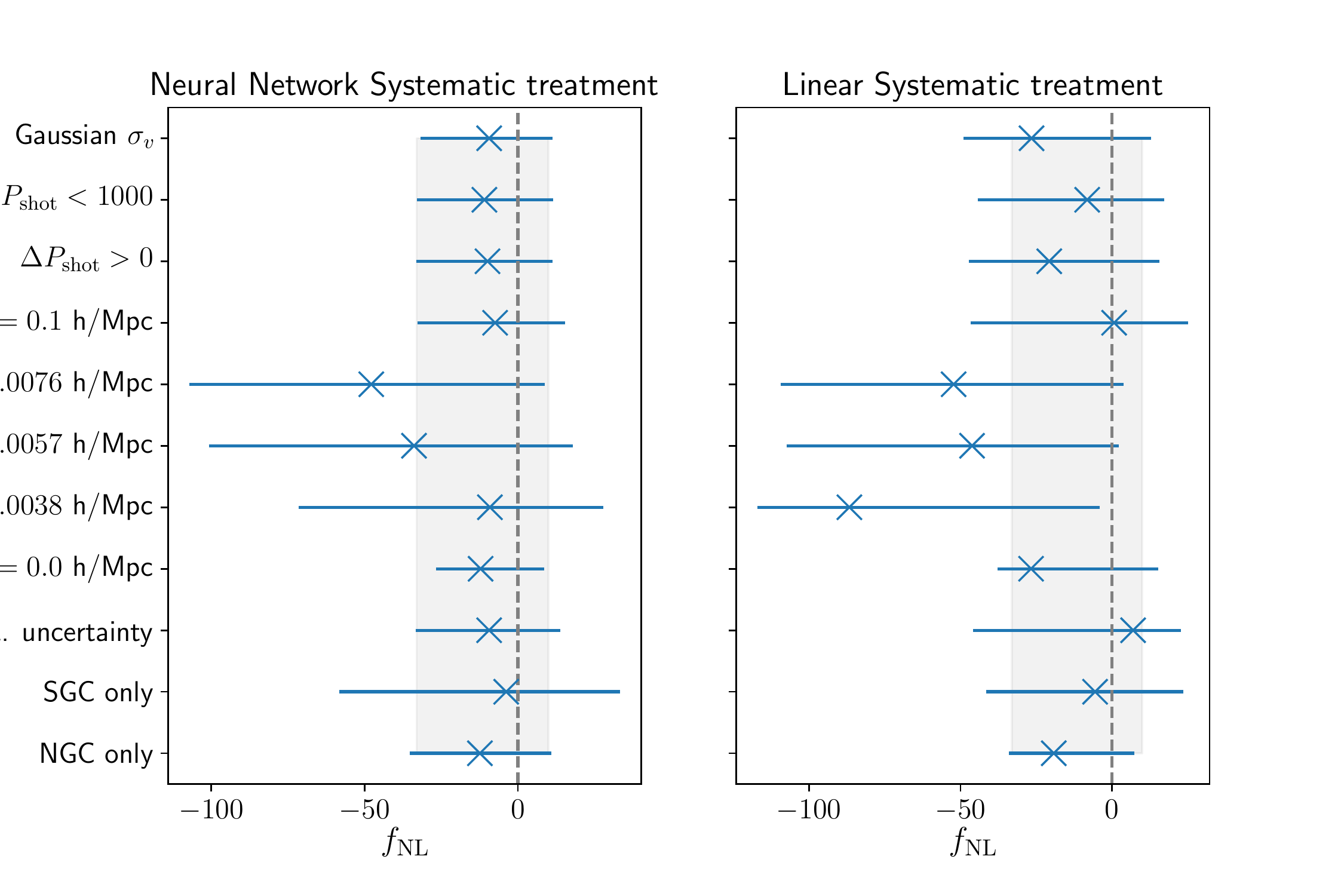}
    \caption{Robustness tests: 68\% C.L constraints for different analysis settings and assumptions on the theoretical model. The Neural Network systematic treatment gives consistent results, while the standard linear systematic removal techniques shows scatter. The grey dashed lines correspond to zero \fNLloc. }
    \label{fig:Robustness}
\end{figure*}

\begin{table*}
	 \centering
    	\begin{tabular}{|l|c|c|c|c|c|c||c|c|c|c|c|}
	\hline
	\hline
	&  \multicolumn{3}{|c|}{Neural Networks Systematic treatment} &  \multicolumn{3}{|c|}{Linear Systematic treatment}  \\
	\hline
	Assumption & \fNL 68\% C.L.& Max. Like. & $\chi^2/\mathrm{dof}$ & \fNL 68\% C.L. & Max. Like. & $\chi^2/\mathrm{dof}$\\
	\hline
	\textbf{Default} & $-33< f_\mathrm{NL}<11$ & -12 & 94/119 & $-50<f_\mathrm{NL}<10$ & -28 & 87/119 \\
	\hline
		NGC only & $-35< f_\mathrm{NL}<10$ & -13 & 44/59& $-34<f_\mathrm{NL}<7$ & -19 & 42/59 \\
		SGC only & $-58< f_\mathrm{NL}<33$ & -4 & 50/59& $-42<f_\mathrm{NL}<23$ & -5 & 45/59 \\
	\hline
	\multirow{2}{7em}{Covariance w/o sys. error}  \\
	  &  $-33<f_\mathrm{NL}<14$ & -9  &  105/119   & $-46<f_\mathrm{NL}<23$ & 6  &  99/54 \\
	\hline
	minimum k [$h$/Mpc]\\
	$k_\mathrm{min}=0.0$ &  $-26<f_\mathrm{NL}<8$ & -14 &  95/121 & $-38<f_\mathrm{NL}<15$ & -27 &  87/121 \\
	$k_\mathrm{min}=0.0038$ &  $-71<f_\mathrm{NL}<27$ & -9  &  92/117 & $-117<f_\mathrm{NL}<-4$ & -3 &  86/117 \\
	$k_\mathrm{min}=0.0057$ &  $-101<f_\mathrm{NL}<17$ & -34  &  91/115& $-108<f_\mathrm{NL}<2$ & -48 &  86/115  \\
	$k_\mathrm{min}=0.0076$  &  $-107<f_\mathrm{NL}<8$ & -48  &  91/113& $-109<f_\mathrm{NL}<4$ & -52 &  86/113 \\
	\hline
	maximum k [$h$/Mpc]\\
	$k_\mathrm{max}=0.1$ &  $-33<f_\mathrm{NL}<15.$ & -8 &  74/99 & $-46<f_\mathrm{NL}<25$ & 1 &  65/99  \\
	%$k_\mathrm{max}=0.112$ & $-92<f_\mathrm{NL}<51$ & -19 &  84/51 &  $-83<f_\mathrm{NL}<53$ & -14  &  92/51 \\
	\hline
	shot noise \\
	$\Delta P_\mathrm{shot}>0 $  &  $-33<f_\mathrm{NL}<11$ & -11  &  95/119& $-47<f_\mathrm{NL}<16$ & -20 &  88/119 \\
	$-1000<\Delta P_\mathrm{shot}>1000$&  $-32<f_\mathrm{NL}<11$ & -9  &  94/119&$-44<f_\mathrm{NL}<17$ & -8& 87/119 \\
	\hline
	Gaussian $\sigma_v$ &  $-33<f_\mathrm{NL}<11$ & -9 &  94/119& $-49<f_\mathrm{NL}<-13$&-27 & 87/119 \\
	\hline
	\end{tabular}
	\caption{Robustness tests using different analysis settings. These results are also displayed in Figure~\ref{fig:Robustness}. The 'Default' values (third row) are the main result of the paper.}
	\label{tab:Robustness}
\end{table*}
 
In this section we investigate the effect of different assumptions in the analysis, for instance, the covariance matrix without systematic contribution, the theoretical model as well as scale cuts and assumptions on priors for both the neural network systematics treatment as well as the linear systematics treatment. Table~\ref{tab:Robustness} and Figure~\ref{fig:Robustness} summarise our results. Overall, the results using the neural network systematics treatment are robust to all assumptions and analysis settings, whereas the linear systematic treatment shows a wider scatter in the resulting \fNL constraints. In the following we discuss each scenario in more detail.

First, we investigate the effect of using the covariance calculated from the `null' mocks, i.e. the purely statistical covariance matrix that does not include any uncertainty due to the systematic treatment, in contrast to our default covariance that includes scatter caused by the systematic weights (see Section~\ref{sec:cov}). In the neural network scenario, the constraints on \fNL are consistent with our default results with only a mild increase in the $\chi^2/$dof, whereas in the linear scenario the maximum likelihood value for \fNL is shifted by more than 1 $\sigma$ showing that the linear results are not robust.

Next, we check the impact of using different scale cuts by changing the minimum and maximum scales fitted, $k_\mathrm{min}$ and $k_\mathrm{max}$. For the neural network technique, increasing $k_\mathrm{min}$ to remove large scales, leads to an increase in the uncertainty in agreement with the $1/k^2$ dependency of the PNG model and only mildly shifts the maximum likelihood to lower values. The linear treatment on the other hand displays strong changes to the \fNL constraints, hinting at residue systematic contamination. Removing the small scale information by decreasing $k_\mathrm{max}$ has a minimal impact on the results for both scenarios, only marginally increasing the uncertainty. 
% confirming that assuming a linear model in the analysis is sufficient. 

Additionally, we show that the prior on the shot noise has no influence on the results for both scenarios. Lastly, we also note that the choice of a Gaussian velocity dispersion in contrast to a Lorentzian behaviour does not change our results. 

\subsection{Potential bias caused by the systematic treatment}

Any treatment of systematics is liable to bias our results towards negative \fNL values by removing clustering modes and thus decreasing the amplitude of the power spectrum monopole on large scales. In Section~\ref{sec:sys} we estimated this effect using contaminated mocks to give $\Delta f_\mathrm{NL}=-12$. Un-biasing our results by shifting the likelihood by this value implies a \fNL constraint of  $f_\mathrm{NL}=0\pm21\mathrm{ (stat.)} \pm 12 \mathrm{ (sys.)}$. This over-correction is significantly less than the statistical error. Since our estimate of the bias relies on mocks with linear systeamtics that do not fully represent the non-linearities of the systematics in the actual data, we do not include this bias in our main results.

%%%%%%%%%%%%%%%%%%% CONCLUSION %%%%%%%%%%%%%%%%%%%
\section{Conclusions}
\label{sec:conclusion}
\begin{figure*}
    \centering
    \includegraphics{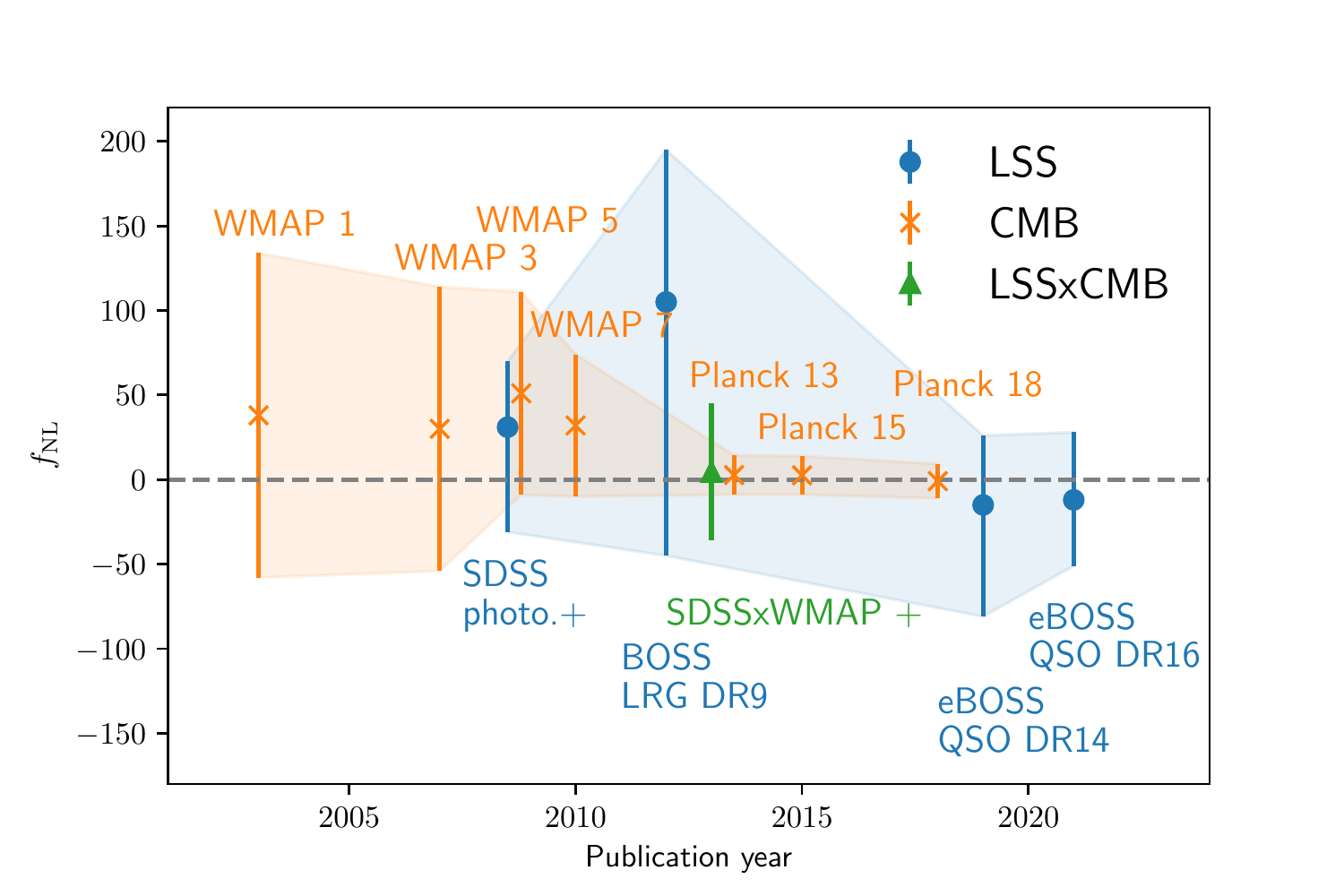}
    \caption{Summary of PNG measurements at 95 \% C.L. from various CMB surveys \protect \citep{Komatsu:2003fd,Spergel_2007,Komatsu_2009,Komatsu_2010,planck13PNG,planck15PNG,planck19PNG}  and LSS data \protect \citep{Slosar08, ross_clustering_2013,eboss-dr14-fnl} including this work with $-51<f_\mathrm{NL}<28$. Note, that we used the median \fNL value in case the best-fit or maximum likelihood value was not quoted in the reference. The grey dashed line corresponds to zero \fNLloc.}
    \label{fig:fNL_history}
\end{figure*}

The quasar sample of eBOSS is the largest volume survey up to date, covering a redshift range of $0.8<z<2.2$. The large survey volume allows the measurement of the power spectrum at very large scales to an unprecedented precision. We measure the amplitude of local PNG to be $f_\mathrm{NL}=-12\pm21$ at 68 $\%$ confidence level finding no evidence for PNG within the 1-$\sigma$ uncertainty. This is the tightest constraint on \fNL from low redshift large-scale structure to date. Figure~\ref{fig:fNL_history} shows this result in context of previous analysis. In orange are the main measurements on \fNL from CMB experiments over the years \citep{Komatsu:2003fd,Spergel_2007,Komatsu_2009,Komatsu_2010,planck13PNG,planck15PNG,planck19PNG}, in blue constraints from LSS surveys \citep{Slosar08, ross_clustering_2013,eboss-dr14-fnl} and in green results from cross-correlations of LSS and CMB \citep{Giannantonio_2014}.
We can improve upon the previous eBOSS DR14 results \citep{eboss-dr14-fnl} by ~35\%, by a factor of 3 upon the BOSS LRG DR9 results \citep{ross_clustering_2013} and by 28\% upon the constrains of \cite{Slosar08}, which used a combination of data sets with the results dominated by the SDSS photometric quasar sample.

We apply optimised weights including redshift weights developed in \cite{Mueller19} as well as FKP weights more suitable for large scales to improve our results by 37\% highlighting the potential of advanced weighting techniques. 

We show that a robust removal of the photometric contamination is crucial for a precise PNG measurement using the power spectrum monopole. Imaging systematics such as stellar density, extinction or depth, can have a immense effect in particular on the largest scales and therefore have the potential to bias the \fNL measurement. 

In addition to the linear weights of the standard catalogue as were used in the BOSS and eBOSS BAO and RSD analysis, we present results using a neural network treatment of systematics as detailed in the companion paper \cite{RezaiCompanion}.  We find robust results using the advanced neural network technique, on the other side, while the linear regression does not give robust results. We conclude, that there are uncorrected contaminants present in the data when using the linear regression technique. Therefore, for a PNG measurement the standard treatment of systematics in not sufficient. 

This paper is the latest step in the chain of development required to make \fNL measurements from galaxy redshift surveys. We have demonstrated an increasingly sophisticated pipeline including careful choice of weights and analysis of potential problems caused by the observations. These developments are particularly important given future surveys including DESI \citep{desi_collaboration_desi_2016,desi_collaboration_desi_2016-1}, SPHEREx \citep{Dore-spherex-cosmo} and Euclid \citep{laureijs_euclid_2011}, which will cover larger effective volumes than that considered here. Dedicated experiments such as a survey of high redshift galaxies by an instrument such as the Maunakea Spectroscopic Explorer \citep{MSE-fnl} offer a potential to achieve errors of order $\Delta f_\mathrm{NL}\sim1$, which is the level at which interesting constraints can be set on inflation \citep{Bartolo04}. Robsutly realising this potential will require further developments, particularly for the mitigation of systematics arising from the observations. Including measurements of the bispectrum and trispectrum have the potential to significantly improve on these constraints \citep{Karagiannis18} further increasing the impact of large-scale structure observations. What is clear is that over the next decade, large-scale structure has the potential overtake the CMB as the primary observation used to constrain \fNL.

\section*{Acknowledgements}
E.-M. M. would like to thank Chris Blake for helpful discussions.
E.-M. M. acknowledges support from the European Research Council (ERC) under the European Union’s Horizon 2020 research and innovation programme (grant agreement No 693024). M.R. and H.-J.S. are supported by the U.S.~Department of Energy, Office of Science, Office of High Energy Physics under DE-SC0014329 and DE-SC0019091. 
This work made substantial use of open-source software and modules, such as Nbodykit, \textsc{HEALPix}, Fitsio, NumPy, SciPy, IPython, Jupyter, and GitHub.

Funding for the Sloan Digital Sky 
Survey IV has been provided by the 
Alfred P. Sloan Foundation, the U.S. 
Department of Energy Office of 
Science, and the Participating 
Institutions. 

SDSS-IV acknowledges support and 
resources from the Center for High 
Performance Computing  at the 
University of Utah. The SDSS 
website is www.sdss.org.

SDSS-IV is managed by the 
Astrophysical Research Consortium 
for the Participating Institutions 
of the SDSS Collaboration including 
the Brazilian Participation Group, 
the Carnegie Institution for Science, 
Carnegie Mellon University, Center for 
Astrophysics | Harvard \& 
Smithsonian, the Chilean Participation 
Group, the French Participation Group, 
Instituto de Astrof\'isica de 
Canarias, The Johns Hopkins 
University, Kavli Institute for the 
Physics and Mathematics of the 
Universe (IPMU) / University of 
Tokyo, the Korean Participation Group, 
Lawrence Berkeley National Laboratory, 
Leibniz Institut f\"ur Astrophysik 
Potsdam (AIP),  Max-Planck-Institut 
f\"ur Astronomie (MPIA Heidelberg), 
Max-Planck-Institut f\"ur 
Astrophysik (MPA Garching), 
Max-Planck-Institut f\"ur 
Extraterrestrische Physik (MPE), 
National Astronomical Observatories of 
China, New Mexico State University, 
New York University, University of 
Notre Dame, Observat\'ario 
Nacional / MCTI, The Ohio State 
University, Pennsylvania State 
University, Shanghai 
Astronomical Observatory, United 
Kingdom Participation Group, 
Universidad Nacional Aut\'onoma 
de M\'exico, University of Arizona, 
University of Colorado Boulder, 
University of Oxford, University of 
Portsmouth, University of Utah, 
University of Virginia, University 
of Washington, University of 
Wisconsin, Vanderbilt University, 
and Yale University.

\section*{DATA AVAILABILITY}
The new catalogue used in this article will be made public in Github at \url{https://github.com/mehdirezaie/eBOSSDR16QSOE}. The details of the neural network catalog creation are described in \cite{RezaiCompanion}.
The standard LSS catalog used in this work is available at \url{https://www.sdss.org/dr16/spectro/lss/}.
%%%%%%%%%%%%%%%%%%%%%%%%%%%%%%%%%%%%%%%%%%%%%%%%%%

%%%%%%%%%%%%%%%%%%%% REFERENCES %%%%%%%%%%%%%%%%%%

% The best way to enter references is to use BibTeX:

\bibliographystyle{mnras}
\bibliography{eBOSSfNL.bib} % if your bibtex file is called example.bib

\begin{thebibliography}{}
\makeatletter
\relax
\def\mn@urlcharsother{\let\do\@makeother \do\$\do\&\do\#\do\^\do\_\do\%\do\~}
\def\mn@doi{\begingroup\mn@urlcharsother \@ifnextchar [ {\mn@doi@}
  {\mn@doi@[]}}
\def\mn@doi@[#1]#2{\def\@tempa{#1}\ifx\@tempa\@empty \href
  {http://dx.doi.org/#2} {doi:#2}\else \href {http://dx.doi.org/#2} {#1}\fi
  \endgroup}
\def\mn@eprint#1#2{\mn@eprint@#1:#2::\@nil}
\def\mn@eprint@arXiv#1{\href {http://arxiv.org/abs/#1} {{\tt arXiv:#1}}}
\def\mn@eprint@dblp#1{\href {http://dblp.uni-trier.de/rec/bibtex/#1.xml}
  {dblp:#1}}
\def\mn@eprint@#1:#2:#3:#4\@nil{\def\@tempa {#1}\def\@tempb {#2}\def\@tempc
  {#3}\ifx \@tempc \@empty \let \@tempc \@tempb \let \@tempb \@tempa \fi \ifx
  \@tempb \@empty \def\@tempb {arXiv}\fi \@ifundefined
  {mn@eprint@\@tempb}{\@tempb:\@tempc}{\expandafter \expandafter \csname
  mn@eprint@\@tempb\endcsname \expandafter{\@tempc}}}

\bibitem[\protect\citeauthoryear{Ade et~al.,}{Ade et~al.}{2014}]{planck13PNG}
Ade P. A.~R.,  et~al., 2014, \mn@doi [Astronomy & Astrophysics]
  {10.1051/0004-6361/201321554}, 571, A24

\bibitem[\protect\citeauthoryear{Ade et~al.,}{Ade et~al.}{2016}]{planck15PNG}
Ade P. A.~R.,  et~al., 2016, \mn@doi [Astronomy & Astrophysics]
  {10.1051/0004-6361/201525836}, 594, A17

\bibitem[\protect\citeauthoryear{{Ahumada} et~al.,}{{Ahumada}
  et~al.}{2020}]{SDSS-DR16}
{Ahumada} R.,  et~al., 2020, \mn@doi [\apjs] {10.3847/1538-4365/ab929e}, \href
  {https://ui.adsabs.harvard.edu/abs/2020ApJS..249....3A} {249, 3}

\bibitem[\protect\citeauthoryear{Ata et~al.}{Ata et~al.}{2018}]{Ata:2017dya}
Ata M.,  et~al., 2018, \mn@doi [Mon. Not. Roy. Astron. Soc.]
  {10.1093/mnras/stx2630}, 473, 4773

\bibitem[\protect\citeauthoryear{Bardeen, Bond, Kaiser  \& Szalay}{Bardeen
  et~al.}{1986}]{bardeen_statistics_1986}
Bardeen J.~M.,  Bond J.~R.,  Kaiser N.,   Szalay A.~S.,  1986, \mn@doi [The
  Astrophysical Journal] {10.1086/164143}, 304, 15

\bibitem[\protect\citeauthoryear{{Bartolo}, {Komatsu}, {Matarrese}  \&
  {Riotto}}{{Bartolo} et~al.}{2004}]{Bartolo04}
{Bartolo} N.,  {Komatsu} E.,  {Matarrese} S.,   {Riotto} A.,  2004, \mn@doi
  [\physrep] {10.1016/j.physrep.2004.08.022}, \href
  {https://ui.adsabs.harvard.edu/abs/2004PhR...402..103B} {402, 103}

\bibitem[\protect\citeauthoryear{Bautista et~al.,}{Bautista
  et~al.}{2018}]{bautista2018sdss}
Bautista J.~E.,  et~al., 2018, The Astrophysical Journal, 863, 110

\bibitem[\protect\citeauthoryear{{Bautista} et~al.,}{{Bautista}
  et~al.}{2020}]{LRG_corr}
{Bautista} J.~E.,  et~al., 2020, arXiv e-prints, \href
  {https://ui.adsabs.harvard.edu/abs/2020arXiv200708993B} {p. arXiv:2007.08993}

\bibitem[\protect\citeauthoryear{{Beutler} et~al.,}{{Beutler}
  et~al.}{2017}]{beutler_clustering_2017}
{Beutler} F.,  et~al., 2017, \mn@doi [\mnras] {10.1093/mnras/stw3298}, \href
  {https://ui.adsabs.harvard.edu/abs/2017MNRAS.466.2242B} {466, 2242}

\bibitem[\protect\citeauthoryear{{Bianchi}, {Gil-Mar{\'\i}n}, {Ruggeri}  \&
  {Percival}}{{Bianchi} et~al.}{2015}]{Bianchietal:2015}
{Bianchi} D.,  {Gil-Mar{\'\i}n} H.,  {Ruggeri} R.,   {Percival} W.~J.,  2015,
  \mn@doi [\mnras] {10.1093/mnrasl/slv090}, \href
  {https://ui.adsabs.harvard.edu/abs/2015MNRAS.453L..11B} {453, L11}

\bibitem[\protect\citeauthoryear{Blanton et~al.,}{Blanton
  et~al.}{2017}]{blanton_sloan_2017}
Blanton M.~R.,  et~al., 2017, \mn@doi [The Astronomical Journal]
  {10.3847/1538-3881/aa7567}, 154, 28

\bibitem[\protect\citeauthoryear{{Castorina} et~al.,}{{Castorina}
  et~al.}{2019}]{eboss-dr14-fnl}
{Castorina} E.,  et~al., 2019, \mn@doi [\jcap] {10.1088/1475-7516/2019/09/010},
  \href {https://ui.adsabs.harvard.edu/abs/2019JCAP...09..010C} {2019, 010}

\bibitem[\protect\citeauthoryear{Chuang, Kitaura, Prada, Zhao  \& Yepes}{Chuang
  et~al.}{2014}]{Chuang_2014}
Chuang C.-H.,  Kitaura F.-S.,  Prada F.,  Zhao C.,   Yepes G.,  2014, \mn@doi
  [Monthly Notices of the Royal Astronomical Society] {10.1093/mnras/stu2301},
  446, 2621–2628

\bibitem[\protect\citeauthoryear{Collaboration et~al.,}{Collaboration
  et~al.}{2019}]{planck19PNG}
Collaboration P.,  et~al., 2019, Planck 2018 results. IX. Constraints on
  primordial non-Gaussianity (\mn@eprint {arXiv} {1905.05697})

\bibitem[\protect\citeauthoryear{Crocce et~al.,}{Crocce
  et~al.}{2015}]{Crocce2015}
Crocce M.,  et~al., 2015, \mn@doi [Monthly Notices of the Royal Astronomical
  Society] {10.1093/mnras/stv2590}, 455, 4301

\bibitem[\protect\citeauthoryear{{DESI Collaboration} et~al.,}{{DESI
  Collaboration} et~al.}{2016a}]{desi_collaboration_desi_2016}
{DESI Collaboration} et~al., 2016a, arXiv e-prints, \href
  {https://ui.adsabs.harvard.edu/abs/2016arXiv161100036D} {p. arXiv:1611.00036}

\bibitem[\protect\citeauthoryear{{DESI Collaboration} et~al.,}{{DESI
  Collaboration} et~al.}{2016b}]{desi_collaboration_desi_2016-1}
{DESI Collaboration} et~al., 2016b, arXiv e-prints, \href
  {https://ui.adsabs.harvard.edu/abs/2016arXiv161100037D} {p. arXiv:1611.00037}

\bibitem[\protect\citeauthoryear{Dalal, Doré, Huterer  \& Shirokov}{Dalal
  et~al.}{2008}]{dalal_imprints_2008}
Dalal N.,  Doré O.,  Huterer D.,   Shirokov A.,  2008, \mn@doi [Physical
  Review D] {10.1103/PhysRevD.77.123514}, 77

\bibitem[\protect\citeauthoryear{Dawson et~al.,}{Dawson
  et~al.}{2016}]{dawson_sdss-iv_2016}
Dawson K.~S.,  et~al., 2016, \mn@doi [The Astronomical Journal]
  {10.3847/0004-6256/151/2/44}, 151, 44

\bibitem[\protect\citeauthoryear{{Dor{\'e}} et~al.,}{{Dor{\'e}}
  et~al.}{2014}]{Dore-spherex-cosmo}
{Dor{\'e}} O.,  et~al., 2014, arXiv e-prints, \href
  {https://ui.adsabs.harvard.edu/abs/2014arXiv1412.4872D} {p. arXiv:1412.4872}

\bibitem[\protect\citeauthoryear{Feldman, Kaiser  \& Peacock}{Feldman
  et~al.}{1994}]{feldman_power-spectrum_1994}
Feldman H.~A.,  Kaiser N.,   Peacock J.~A.,  1994, \mn@doi [The Astrophysical
  Journal] {10.1086/174036}, 426, 23

\bibitem[\protect\citeauthoryear{{Ferramacho}, {Santos}, {Jarvis}  \&
  {Camera}}{{Ferramacho} et~al.}{2014}]{Radio-fnl}
{Ferramacho} L.~D.,  {Santos} M.~G.,  {Jarvis} M.~J.,   {Camera} S.,  2014,
  \mn@doi [\mnras] {10.1093/mnras/stu1015}, \href
  {https://ui.adsabs.harvard.edu/abs/2014MNRAS.442.2511F} {442, 2511}

\bibitem[\protect\citeauthoryear{{Gaia Collaboration} et~al.,}{{Gaia
  Collaboration} et~al.}{2018}]{gaia2018A&A...616A...1G}
{Gaia Collaboration} et~al., 2018, \mn@doi [\aap]
  {10.1051/0004-6361/201833051}, \href
  {https://ui.adsabs.harvard.edu/abs/2018A&A...616A...1G} {616, A1}

\bibitem[\protect\citeauthoryear{{Gangui}, {Lucchin}, {Matarrese}  \&
  {Mollerach}}{{Gangui} et~al.}{1994}]{1994ApJ...430..447G}
{Gangui} A.,  {Lucchin} F.,  {Matarrese} S.,   {Mollerach} S.,  1994, \mn@doi
  [\apj] {10.1086/174421}, \href
  {https://ui.adsabs.harvard.edu/abs/1994ApJ...430..447G} {430, 447}

\bibitem[\protect\citeauthoryear{Giannantonio, Ross, Percival, Crittenden,
  Bacher, Kilbinger, Nichol  \& Weller}{Giannantonio
  et~al.}{2014a}]{Giannantonio_2014}
Giannantonio T.,  Ross A.~J.,  Percival W.~J.,  Crittenden R.,  Bacher D.,
  Kilbinger M.,  Nichol R.,   Weller J.,  2014a, \mn@doi [Physical Review D]
  {10.1103/physrevd.89.023511}, 89

\bibitem[\protect\citeauthoryear{{Giannantonio}, {Ross}, {Percival},
  {Crittenden}, {Bacher}, {Kilbinger}, {Nichol}  \& {Weller}}{{Giannantonio}
  et~al.}{2014b}]{Giannantonio14-fnl}
{Giannantonio} T.,  {Ross} A.~J.,  {Percival} W.~J.,  {Crittenden} R.,
  {Bacher} D.,  {Kilbinger} M.,  {Nichol} R.,   {Weller} J.,  2014b, \mn@doi
  [\prd] {10.1103/PhysRevD.89.023511}, \href
  {https://ui.adsabs.harvard.edu/abs/2014PhRvD..89b3511G} {89, 023511}

\bibitem[\protect\citeauthoryear{{Gil-Mar{\'\i}n} et~al.,}{{Gil-Mar{\'\i}n}
  et~al.}{2020}]{gil-marin20a}
{Gil-Mar{\'\i}n} H.,  et~al., 2020, arXiv e-prints, \href
  {https://ui.adsabs.harvard.edu/abs/2020arXiv200708994G} {p. arXiv:2007.08994}

\bibitem[\protect\citeauthoryear{{G{\'o}rski}, {Hivon}, {Banday}, {Wandelt},
  {Hansen}, {Reinecke}  \& {Bartelmann}}{{G{\'o}rski}
  et~al.}{2005}]{gorski2005ApJ...622..759G}
{G{\'o}rski} K.~M.,  {Hivon} E.,  {Banday} A.~J.,  {Wandelt} B.~D.,  {Hansen}
  F.~K.,  {Reinecke} M.,   {Bartelmann} M.,  2005, \mn@doi [\apj]
  {10.1086/427976}, \href
  {https://ui.adsabs.harvard.edu/abs/2005ApJ...622..759G} {622, 759}

\bibitem[\protect\citeauthoryear{Gualdi, Gil-Marin  \& Verde}{Gualdi
  et~al.}{2021}]{gualdi2021joint}
Gualdi D.,  Gil-Marin H.,   Verde L.,  2021, Joint analysis of anisotropic
  power spectrum, bispectrum and trispectrum: application to N-body simulations
  (\mn@eprint {arXiv} {2104.03976})

\bibitem[\protect\citeauthoryear{{Gunn} \& {Gott}}{{Gunn} \&
  {Gott}}{1972}]{Gunn_Gott_sphericalcollapse}
{Gunn} J.~E.,  {Gott} J.~Richard I.,  1972, \mn@doi [\apj] {10.1086/151605},
  \href {https://ui.adsabs.harvard.edu/abs/1972ApJ...176....1G} {176, 1}

\bibitem[\protect\citeauthoryear{Gunn et~al.}{Gunn et~al.}{2006}]{Gunn:2006tw}
Gunn J.~E.,  et~al., 2006, \mn@doi [Astron. J.] {10.1086/500975}, 131, 2332

\bibitem[\protect\citeauthoryear{{Hartlap}, {Simon}  \& {Schneider}}{{Hartlap}
  et~al.}{2007}]{Hartlap07}
{Hartlap} J.,  {Simon} P.,   {Schneider} P.,  2007, \mn@doi [\aap]
  {10.1051/0004-6361:20066170}, \href
  {https://ui.adsabs.harvard.edu/abs/2007A&A...464..399H} {464, 399}

\bibitem[\protect\citeauthoryear{{Ho} et~al.,}{{Ho} et~al.}{2015}]{Ho15-fnl}
{Ho} S.,  et~al., 2015, \mn@doi [\jcap] {10.1088/1475-7516/2015/05/040}, \href
  {https://ui.adsabs.harvard.edu/abs/2015JCAP...05..040H} {2015, 040}

\bibitem[\protect\citeauthoryear{{Hou} et~al.,}{{Hou}
  et~al.}{2021}]{Hou-eboss-qso-xi}
{Hou} J.,  et~al., 2021, \mn@doi [\mnras] {10.1093/mnras/staa3234}, \href
  {https://ui.adsabs.harvard.edu/abs/2021MNRAS.500.1201H} {500, 1201}

\bibitem[\protect\citeauthoryear{{Kalus}, {Percival}, {Bacon}, {Mueller},
  {Samushia}, {Verde}, {Ross}  \& {Bernal}}{{Kalus} et~al.}{2019}]{Kalus19}
{Kalus} B.,  {Percival} W.~J.,  {Bacon} D.~J.,  {Mueller} E.~M.,  {Samushia}
  L.,  {Verde} L.,  {Ross} A.~J.,   {Bernal} J.~L.,  2019, \mn@doi [\mnras]
  {10.1093/mnras/sty2655}, \href
  {https://ui.adsabs.harvard.edu/abs/2019MNRAS.482..453K} {482, 453}

\bibitem[\protect\citeauthoryear{{Karagiannis}, {Lazanu}, {Liguori},
  {Raccanelli}, {Bartolo}  \& {Verde}}{{Karagiannis}
  et~al.}{2018}]{Karagiannis18}
{Karagiannis} D.,  {Lazanu} A.,  {Liguori} M.,  {Raccanelli} A.,  {Bartolo} N.,
    {Verde} L.,  2018, \mn@doi [\mnras] {10.1093/mnras/sty1029}, \href
  {https://ui.adsabs.harvard.edu/abs/2018MNRAS.478.1341K} {478, 1341}

\bibitem[\protect\citeauthoryear{Komatsu}{Komatsu}{2010}]{Komatsu_2010}
Komatsu E.,  2010, \mn@doi [Classical and Quantum Gravity]
  {10.1088/0264-9381/27/12/124010}, 27, 124010

\bibitem[\protect\citeauthoryear{{Komatsu} \& {Spergel}}{{Komatsu} \&
  {Spergel}}{2001}]{2001PhRvD..63f3002K}
{Komatsu} E.,  {Spergel} D.~N.,  2001, \mn@doi [\prd]
  {10.1103/PhysRevD.63.063002}, \href
  {https://ui.adsabs.harvard.edu/abs/2001PhRvD..63f3002K} {63, 063002}

\bibitem[\protect\citeauthoryear{Komatsu et~al.}{Komatsu
  et~al.}{2003}]{Komatsu:2003fd}
Komatsu E.,  et~al., 2003, \mn@doi [Astrophys. J. Suppl.] {10.1086/377220},
  148, 119

\bibitem[\protect\citeauthoryear{Komatsu et~al.,}{Komatsu
  et~al.}{2009}]{Komatsu_2009}
Komatsu E.,  et~al., 2009, \mn@doi [The Astrophysical Journal Supplement
  Series] {10.1088/0067-0049/180/2/330}, 180, 330–376

\bibitem[\protect\citeauthoryear{{Laureijs} et~al.,}{{Laureijs}
  et~al.}{2011}]{laureijs_euclid_2011}
{Laureijs} R.,  et~al., 2011, arXiv e-prints, \href
  {https://ui.adsabs.harvard.edu/abs/2011arXiv1110.3193L} {p. arXiv:1110.3193}

\bibitem[\protect\citeauthoryear{{Laurent} et~al.,}{{Laurent}
  et~al.}{2017}]{Laurent-dr14-qso}
{Laurent} P.,  et~al., 2017, \mn@doi [\jcap] {10.1088/1475-7516/2017/07/017},
  \href {https://ui.adsabs.harvard.edu/abs/2017JCAP...07..017L} {2017, 017}

\bibitem[\protect\citeauthoryear{{Leistedt}, {Peiris}  \& {Roth}}{{Leistedt}
  et~al.}{2014}]{Leistedt14-fnl}
{Leistedt} B.,  {Peiris} H.~V.,   {Roth} N.,  2014, \mn@doi [\prl]
  {10.1103/PhysRevLett.113.221301}, \href
  {https://ui.adsabs.harvard.edu/abs/2014PhRvL.113v1301L} {113, 221301}

\bibitem[\protect\citeauthoryear{{Lyke} et~al.,}{{Lyke}
  et~al.}{2020}]{Lyke-dr16-qsocat}
{Lyke} B.~W.,  et~al., 2020, \mn@doi [\apjs] {10.3847/1538-4365/aba623}, \href
  {https://ui.adsabs.harvard.edu/abs/2020ApJS..250....8L} {250, 8}

\bibitem[\protect\citeauthoryear{{Matarrese} \& {Verde}}{{Matarrese} \&
  {Verde}}{2008}]{Matarrese08}
{Matarrese} S.,  {Verde} L.,  2008, \mn@doi [\apjl] {10.1086/587840}, \href
  {https://ui.adsabs.harvard.edu/abs/2008ApJ...677L..77M} {677, L77}

\bibitem[\protect\citeauthoryear{{Merz} et~al.,}{{Merz}
  et~al.}{2021}]{Merz2021}
{Merz} G.,  et~al., 2021, arXiv e-prints, \href
  {https://ui.adsabs.harvard.edu/abs/2021arXiv210510463M} {p. arXiv:2105.10463}

\bibitem[\protect\citeauthoryear{Mueller, Percival, Linder, Alam, Zhao,
  Sánchez, Beutler  \& Brinkmann}{Mueller
  et~al.}{2018}]{mueller_clustering_2018}
Mueller E.-M.,  Percival W.,  Linder E.,  Alam S.,  Zhao G.-B.,  Sánchez
  A.~G.,  Beutler F.,   Brinkmann J.,  2018, \mn@doi [Monthly Notices of the
  Royal Astronomical Society] {10.1093/mnras/stx3232}, 475, 2122

\bibitem[\protect\citeauthoryear{{Mueller}, {Percival}  \& {Ruggeri}}{{Mueller}
  et~al.}{2019}]{Mueller19}
{Mueller} E.-M.,  {Percival} W.~J.,   {Ruggeri} R.,  2019, \mn@doi [\mnras]
  {10.1093/mnras/sty3150}, \href
  {https://ui.adsabs.harvard.edu/abs/2019MNRAS.485.4160M} {485, 4160}

\bibitem[\protect\citeauthoryear{Myers et~al.}{Myers
  et~al.}{2015}]{Myers:2015hpw}
Myers A.~D.,  et~al., 2015, \mn@doi [Astrophys. J. Suppl.]
  {10.1088/0067-0049/221/2/27}, 221, 27

\bibitem[\protect\citeauthoryear{{Neveux} et~al.,}{{Neveux}
  et~al.}{2020}]{Neveux-eboss-qso-pk}
{Neveux} R.,  et~al., 2020, \mn@doi [\mnras] {10.1093/mnras/staa2780}, \href
  {https://ui.adsabs.harvard.edu/abs/2020MNRAS.499..210N} {499, 210}

\bibitem[\protect\citeauthoryear{{Percival} et~al.,}{{Percival}
  et~al.}{2019}]{MSE-fnl}
{Percival} W.~J.,  et~al., 2019, arXiv e-prints, \href
  {https://ui.adsabs.harvard.edu/abs/2019arXiv190303158P} {p. arXiv:1903.03158}

\bibitem[\protect\citeauthoryear{{Planck Collaboration} et~al.,}{{Planck
  Collaboration} et~al.}{2020}]{Planck2018-fnl}
{Planck Collaboration} et~al., 2020, \mn@doi [\aap]
  {10.1051/0004-6361/201935891}, \href
  {https://ui.adsabs.harvard.edu/abs/2020A&A...641A...9P} {641, A9}

\bibitem[\protect\citeauthoryear{{Raichoor} et~al.,}{{Raichoor}
  et~al.}{2020}]{raichoor20a}
{Raichoor} A.,  et~al., 2020, arXiv e-prints, \href
  {https://ui.adsabs.harvard.edu/abs/2020arXiv200709007R} {p. arXiv:2007.09007}

\bibitem[\protect\citeauthoryear{Rezaie et~al.,}{Rezaie
  et~al.}{2021}]{RezaiCompanion}
Rezaie M.,  et~al., 2021, \mn@doi [Monthly Notices of the Royal Astronomical
  Society] {10.1093/mnras/stab1730}

\bibitem[\protect\citeauthoryear{Ross et~al.,}{Ross
  et~al.}{2013b}]{ross_clustering_2013}
Ross A.~J.,  et~al., 2013b, \mn@doi [Monthly Notices of the Royal Astronomical
  Society] {10.1093/mnras/sts094}, 428, 1116

\bibitem[\protect\citeauthoryear{{Ross} et~al.,}{{Ross}
  et~al.}{2013a}]{Ross13-fnl}
{Ross} A.~J.,  et~al., 2013a, \mn@doi [\mnras] {10.1093/mnras/sts094}, \href
  {https://ui.adsabs.harvard.edu/abs/2013MNRAS.428.1116R} {428, 1116}

\bibitem[\protect\citeauthoryear{Ross, Bautista, Tojeiro, Brownstein, Burtin,
  Dawson  \& et al.}{Ross et~al.}{2020a}]{ebossDR16catalogue}
Ross A.~J.,  Bautista J.,  Tojeiro R.,  Brownstein J.~R.,  Burtin E.,  Dawson
  K.~S.,   et al. 2020a, \mn@doi [\mnras] {10.1093/mnras/stv0000}, 000, 000

\bibitem[\protect\citeauthoryear{{Ross} et~al.,}{{Ross}
  et~al.}{2020b}]{Rosss2020MNRAS.498.2354R}
{Ross} A.~J.,  et~al., 2020b, \mn@doi [\mnras] {10.1093/mnras/staa2416}, \href
  {https://ui.adsabs.harvard.edu/abs/2020MNRAS.498.2354R} {498, 2354}

\bibitem[\protect\citeauthoryear{Ruggeri, Percival, Gil-Marín, Zhu, Zhao  \&
  Wang}{Ruggeri et~al.}{2017}]{ruggeri_optimal_2017}
Ruggeri R.,  Percival W.~J.,  Gil-Marín H.,  Zhu F.,  Zhao G.-B.,   Wang Y.,
  2017, \mn@doi [Monthly Notices of the Royal Astronomical Society]
  {10.1093/mnras/stw2422}, 464, 2698

\bibitem[\protect\citeauthoryear{Schlegel, Finkbeiner  \& Davis}{Schlegel
  et~al.}{1998a}]{Schlegel:1997yv}
Schlegel D.~J.,  Finkbeiner D.~P.,   Davis M.,  1998a, \mn@doi [Astrophys. J.]
  {10.1086/305772}, 500, 525

\bibitem[\protect\citeauthoryear{Schlegel, Finkbeiner  \& Davis}{Schlegel
  et~al.}{1998b}]{schlegel1998maps}
Schlegel D.~J.,  Finkbeiner D.~P.,   Davis M.,  1998b, The Astrophysical
  Journal, 500, 525

\bibitem[\protect\citeauthoryear{{Scoccimarro}}{{Scoccimarro}}{2015}]{Soccimarro:2015}
{Scoccimarro} R.,  2015, \mn@doi [\prd] {10.1103/PhysRevD.92.083532}, \href
  {https://ui.adsabs.harvard.edu/abs/2015PhRvD..92h3532S} {92, 083532}

\bibitem[\protect\citeauthoryear{{Sefusatti}, {Crocce}, {Scoccimarro}  \&
  {Couchman}}{{Sefusatti} et~al.}{2016}]{Sefusattietal:2016}
{Sefusatti} E.,  {Crocce} M.,  {Scoccimarro} R.,   {Couchman} H.~M.~P.,  2016,
  \mn@doi [\mnras] {10.1093/mnras/stw1229}, \href
  {https://ui.adsabs.harvard.edu/abs/2016MNRAS.460.3624S} {460, 3624}

\bibitem[\protect\citeauthoryear{{Slosar}, {Hirata}, {Seljak}, {Ho}  \&
  {Padmanabhan}}{{Slosar} et~al.}{2008}]{Slosar08}
{Slosar} A.,  {Hirata} C.,  {Seljak} U.,  {Ho} S.,   {Padmanabhan} N.,  2008,
  \mn@doi [\jcap] {10.1088/1475-7516/2008/08/031}, \href
  {https://ui.adsabs.harvard.edu/abs/2008JCAP...08..031S} {2008, 031}

\bibitem[\protect\citeauthoryear{Smee et~al.}{Smee et~al.}{2013}]{Smee:2012wd}
Smee S.,  et~al., 2013, \mn@doi [Astron. J.] {10.1088/0004-6256/146/2/32}, 146,
  32

\bibitem[\protect\citeauthoryear{{Smith} et~al.,}{{Smith}
  et~al.}{2020}]{Smith-eboss-qso-mocks}
{Smith} A.,  et~al., 2020, \mn@doi [\mnras] {10.1093/mnras/staa2825}, \href
  {https://ui.adsabs.harvard.edu/abs/2020MNRAS.499..269S} {499, 269}

\bibitem[\protect\citeauthoryear{Spergel et~al.,}{Spergel
  et~al.}{2007}]{Spergel_2007}
Spergel D.~N.,  et~al., 2007, \mn@doi [The Astrophysical Journal Supplement
  Series] {10.1086/513700}, 170, 377–408

\bibitem[\protect\citeauthoryear{{Tamone} et~al.,}{{Tamone}
  et~al.}{2020}]{tamone20a}
{Tamone} A.,  et~al., 2020, arXiv e-prints, \href
  {https://ui.adsabs.harvard.edu/abs/2020arXiv200709009T} {p. arXiv:2007.09009}

\bibitem[\protect\citeauthoryear{{Wang}, {Percival}, {Avila}, {Crittenden}  \&
  {Bianchi}}{{Wang} et~al.}{2019}]{wang19}
{Wang} M.~S.,  {Percival} W.~J.,  {Avila} S.,  {Crittenden} R.,   {Bianchi} D.,
   2019, \mn@doi [\mnras] {10.1093/mnras/stz829}, \href
  {https://ui.adsabs.harvard.edu/abs/2019MNRAS.486..951W} {486, 951}

\bibitem[\protect\citeauthoryear{{Wang}, {Beutler}  \& {Bacon}}{{Wang}
  et~al.}{2020}]{wang20}
{Wang} M.~S.,  {Beutler} F.,   {Bacon} D.,  2020, \mn@doi [\mnras]
  {10.1093/mnras/staa2998}, \href
  {https://ui.adsabs.harvard.edu/abs/2020MNRAS.499.2598W} {499, 2598}

\bibitem[\protect\citeauthoryear{{Wilson}, {Peacock}, {Taylor}  \& {de la
  Torre}}{{Wilson} et~al.}{2017}]{wilson_good_2016}
{Wilson} M.~J.,  {Peacock} J.~A.,  {Taylor} A.~N.,   {de la Torre} S.,  2017,
  \mn@doi [\mnras] {10.1093/mnras/stw2576}, \href
  {https://ui.adsabs.harvard.edu/abs/2017MNRAS.464.3121W} {464, 3121}

\bibitem[\protect\citeauthoryear{{Yamauchi}, {Takahashi}  \&
  {Oguri}}{{Yamauchi} et~al.}{2014}]{Euclid-SKA-fnl}
{Yamauchi} D.,  {Takahashi} K.,   {Oguri} M.,  2014, \mn@doi [\prd]
  {10.1103/PhysRevD.90.083520}, \href
  {https://ui.adsabs.harvard.edu/abs/2014PhRvD..90h3520Y} {90, 083520}

\bibitem[\protect\citeauthoryear{Zhao et~al.,}{Zhao
  et~al.}{2016}]{zhao_extended_2016}
Zhao G.-B.,  et~al., 2016, \mn@doi [Monthly Notices of the Royal Astronomical
  Society] {10.1093/mnras/stw135}, 457, 2377

\bibitem[\protect\citeauthoryear{{Zhao} et~al.,}{{Zhao}
  et~al.}{2019}]{Zhao-rsd-optimal-2019}
{Zhao} G.-B.,  et~al., 2019, \mn@doi [\mnras] {10.1093/mnras/sty2845}, \href
  {https://ui.adsabs.harvard.edu/abs/2019MNRAS.482.3497Z} {482, 3497}

\bibitem[\protect\citeauthoryear{Zhao et~al.,}{Zhao et~al.}{2021}]{Zhao_2021}
Zhao C.,  et~al., 2021, \mn@doi [Monthly Notices of the Royal Astronomical
  Society] {10.1093/mnras/stab510}, 503, 1149–1173

\bibitem[\protect\citeauthoryear{Zhu, Padmanabhan  \& White}{Zhu
  et~al.}{2015}]{zhu_optimal_2015}
Zhu F.,  Padmanabhan N.,   White M.,  2015, \mn@doi [Monthly Notices of the
  Royal Astronomical Society] {10.1093/mnras/stv964}, 451, 236

\bibitem[\protect\citeauthoryear{Zhu, Padmanabhan, White, Ross  \& Zhao}{Zhu
  et~al.}{2016}]{zhu_redshift_2016}
Zhu F.,  Padmanabhan N.,  White M.,  Ross A.~J.,   Zhao G.,  2016, \mn@doi
  [Monthly Notices of the Royal Astronomical Society] {10.1093/mnras/stw1515},
  461, 2867

\bibitem[\protect\citeauthoryear{de Mattia \& Ruhlmann-Kleider}{de~Mattia \&
  Ruhlmann-Kleider}{2019}]{de_Mattia_2019}
de Mattia A.,  Ruhlmann-Kleider V.,  2019, \mn@doi [Journal of Cosmology and
  Astroparticle Physics] {10.1088/1475-7516/2019/08/036}, 2019, 036–036

\bibitem[\protect\citeauthoryear{{de Mattia} et~al.,}{{de Mattia}
  et~al.}{2020}]{demattia20a}
{de Mattia} A.,  et~al., 2020, arXiv e-prints, \href
  {https://ui.adsabs.harvard.edu/abs/2020arXiv200709008D} {p. arXiv:2007.09008}

\bibitem[\protect\citeauthoryear{{du Mas des Bourboux} et~al.,}{{du Mas des
  Bourboux} et~al.}{2020}]{eBOSS-DR16-Lya}
{du Mas des Bourboux} H.,  et~al., 2020, \mn@doi [\apj]
  {10.3847/1538-4357/abb085}, \href
  {https://ui.adsabs.harvard.edu/abs/2020ApJ...901..153D} {901, 153}

\bibitem[\protect\citeauthoryear{{eBOSS Collaboration} et~al.,}{{eBOSS
  Collaboration} et~al.}{2020}]{eBOSS_Cosmology}
{eBOSS Collaboration} et~al., 2020, arXiv e-prints, \href
  {https://ui.adsabs.harvard.edu/abs/2020arXiv200708991E} {p. arXiv:2007.08991}

\makeatother
\end{thebibliography}

% Alternatively you could enter them by hand, like this:
% This method is tedious and prone to error if you have lots of references

%%%%%%%%%%%%%%%%%%%%%%%%%%%%%%%%%%%%%%%%%%%%%%%%%%

%%%%%%%%%%%%%%%%%% APPENDICES %%%%%%%%%%%%%%%%%%%%%
%
%\appendix
%
%\section{Some extra material}
%\begin{figure}
%    \centering
%    \includegraphics[width=0.45\textwidth]{plots/window.png}
%    \caption{Window function multipoles used in this analysis for the NGC (solid lines) and SGC (dashed lines). Higher order multipoles beyond $l>4$ have negligible effect on the convolved power spectrum.}
%    \label{fig:window}
%\end{figure}
%

%If you want to present additional material which would interrupt the flow of the main paper,
%it can be placed in an Appendix which appears after the list of references.
%
%%%%%%%%%%%%%%%%%%%%%%%%%%%%%%%%%%%%%%%%%%%%%%%%%%%
%

% Don't change these lines
\bsp	% typesetting comment
\label{lastpage}
\end{document}